\documentclass[aps,prl,showpacs,floatfix,twocolumn,superscriptaddress]{revtex4-1}
\usepackage{graphicx}
\usepackage{amsmath}
\usepackage{bm}
\usepackage{color}
\usepackage{dcolumn}
\usepackage{multirow}
\usepackage{hyperref}

\begin{document}

\title{Quadruply-ionized barium as a candidate for a high-accuracy optical clock}

\newcommand{\NIST}{
National Institute of Standards and Technology, Boulder, Colorado, USA}

\newcommand{\CSU}{
Department of Physics, Colorado State University, Fort Collins, Colorado, USA}

\newcommand{\UNSW}{
School of Physics, University of New South Wales, Sydney, Australia}

\author{K. Beloy}
\email{kyle.beloy@nist.gov}
\affiliation{\NIST}

\author{V. A. Dzuba}
\affiliation{\UNSW}

\author{S. M. Brewer}
\affiliation{\CSU}

\date{\today}

\begin{abstract}
We identify Ba$^{4+}$ (Te-like) as a promising candidate for a high-accuracy optical clock. The lowest-lying electronic states are part of a $^3P_J$ fine structure manifold with anomalous energy ordering, being non-monotonic in $J$. We propose a clock based on the 338.8 THz electric quadrupole transition between the ground ($^3P_2$) and first-excited ($^3P_0$) electronic states. We perform relativistic many-body calculations to determine relevant properties of this ion. The lifetime of the excited clock state is found to be several seconds, accommodating low statistical uncertainty with a single ion for practical averaging times. The differential static scalar polarizability is found to be small and negative, providing suppressed sensitivity to blackbody radiation while simultaneously allowing cancellation of Stark and excess micromotion shifts. With the exception of Hg$^+$ and Yb$^+$, sensitivity to variation of the fine structure constant is greater than other optical clocks thus far demonstrated.  
\end{abstract}


\maketitle

\newcommand{\later}{\ensuremath{\spadesuit}}

The past two decades have witnessed the rise of atomic clocks based on optical transitions \cite{LudBoyYe15}. State-of-the-art optical atomic clocks interrogate either an ensemble of neutral atoms confined in an optical lattice~\cite{UshTakDas15,NemOhkTak16,McGZhaFas18,BotKedOel19,PizBreBarinpress} or an individual ion confined in a Paul trap~\cite{MadDubZho12,HunSanLip16,HuaGuaBia17,BayGodJon18,BreCheHan19}. In the case of the Al$^+$ clock, with the lowest reported fractional inaccuracy at $9.4\times10^{-19}$~\cite{BreCheHan19}, the single Al$^+$ ``clock'' ion is co-trapped with an ancillary ``logic'' ion (e.g., Be$^+$, Mg$^+$, or Ca$^+$), which facilitates cooling, state initialization, and state detection via quantum-logic spectroscopy~\cite{SchRosLan05}. Beyond their metrological role as frequency references, optical clocks can be used to examine fundamental aspects of nature~\cite{SafBudDeM18} and hold promise for novel applications such as relativistic geodesy~\cite{MehGroLis18,McGZhaFas18}.

Ion clocks demonstrated thus far employ singly-charged ions. Schiller~\cite{Sch07} and Berengut~{\it et al.}~\cite{BerDzuFla10} pointed out that optical transitions in highly-charged ions can amplify signatures of ``new physics,'' such as variation of the fine structure constant $\alpha$, while also possessing a high natural quality factor and insusceptibility to environmental perturbations (generally qualifying them as clock transitions). This spawned a plethora of theoretical works focused on highly-charged ion optical clocks~\cite{BerDzuFla11a,BerDzuFla11b,BerDzuFla12a,BerDzuFla12b,
DerDzuFla12,DzuDerFla12a,DzuDerFla12b,KozDzuFla13,SafDzuFla14a,SafDzuFla14b,SafDzuFla14c,
YudTaiDer14,DzuFlaKat15,DzuSafSaf15,DzuFla15,KozSafCre18,YuSah19}. 
While several challenges must be overcome to ultimately realize a competitive clock based on highly-charged ions, experimental headway has been made towards this goal~\cite{BreGuiTan13,GuiTanBre14,WinCreBek15,SchVerSch15,LeoKinMic19,BekBorHar19,MicLeoKin20}. In particular, high-resolution quantum logic spectroscopy was recently demonstrated with an Ar$^{13+}$/Be$^+$ ion pair~\cite{MicLeoKin20}. This represents an important step, as highly-charged ions presumably require quantum logic techniques due to an absence of accessible electric dipole ($E1$) cycling transitions. For this proof-of-principle work, a fine structure transition in Ar$^{13+}$ served the role of the clock transition.

Prior to the experimental work of Ref.~\cite{MicLeoKin20}, Yudin {\it et al.}~\cite{YudTaiDer14} proposed optical clocks based on fine structure transitions in multiply-charged ions (with ionization degree between two and 17). They argued that such clocks could potentially realize systematic uncertainties below the $10^{-20}$ fractional level. In addition to this, favorable aspects include theoretical tractability of the atomic structure and low multiplicity of the clock states. The latter is an attractive feature, as fidelity of the quantum logic techniques can be challenged with high multiplicity of the clock states. Despite the benefits of the fine structure transitions proposed in Ref.~\cite{YudTaiDer14}, a notable drawback is that the natural lifetime of the excited clock state limits the statistical uncertainty that can be attained with practical averaging times. Assuming Ramsey spectroscopy with near-unity duty cycle, ideal $\pi/2$ pulse areas, and measurement noise only due to projection of the quantum superposition state~\cite{ItaBerBol93}, the optimal Ramsey interrogation time is equal to the natural lifetime of the excited clock state, with a single-ion fractional instability given by~\cite{PeiSchTam05,KozSafCre18}
\begin{equation}
\sigma_y(t)=\frac{0.412}{\nu_0\sqrt{\tau t}}.
\label{Eq:stabform}
\end{equation}
Here $\nu_0$ is the clock frequency, $\tau$ is the lifetime of the excited clock state, and $t$ is the averaging time. Using Eq.~(\ref{Eq:stabform}), the fine structure transitions considered in Ref.~\cite{YudTaiDer14} all have $\sigma_y(t)>3\times10^{-15}/\sqrt{t/\mathrm{s}}$. An entire year of continuous averaging, for example, would procure a fractional statistical uncertainty no better than $5\times10^{-19}$. By comparison, optical lattice clocks are presently capable of such measurement precision with just a few hours of averaging~\cite{SchBroMcG17,OelHutKen19}.

While long lifetimes are a critical feature for clock states, exceptionally long lifetimes can introduce several challenges. For one, determining the transition frequency adequately enough to undertake precision spectroscopy can be a formidable challenge in itself (an example being the elusive thorium nuclear clock transition~\cite{PeiTam03,SeiWenBil19}). A more enduring problem is posed by the high laser intensity required to directly drive the transition, which may be practically infeasible or lead to large frequency shifts. While methods have been developed to mitigate probe-related shifts in atomic clocks~\cite{YudTaiOat10,SanHunLan18}, even small imperfections in their implementation may translate to appreciable clock error~\cite{Bel18}. Finally, while longer lifetimes imply lower instability according to Eq.~(\ref{Eq:stabform}), local oscillator noise sets a practical constraint on the interrogation time. State-of-the-art cavity-stabilized lasers, developed for optical clocks, presently accommodate interrogation times of tens of seconds~\cite{MatLegHaf17,RobOelMil19}, with no gain to be had from clock state lifetimes greatly exceeding this technical limitation.

\begin{figure}[tb]
\includegraphics[width=246pt]{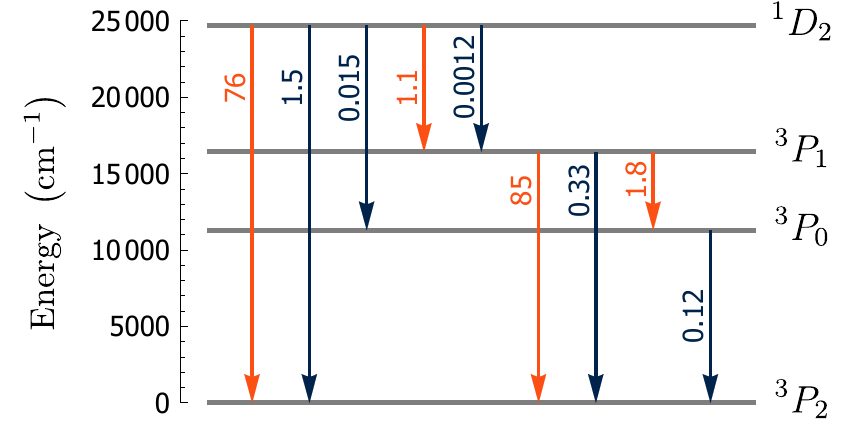}
\caption{The lowest-lying states in Ba$^{4+}$. All states have electronic configuration [Kr]$4d^{10}5s^25p^4$. Arrows show all $M1$ (light orange) and $E2$ (dark blue) decay channels for these states, with calculated decay rates given in units of s$^{-1}$. The proposed clock transition is between the ground state and the first-excited state, ${^3P_2}\rightarrow{^3P_0}$, with a transition frequency of 338.8 THz (wavelength of 884.8 nm) and a natural lifetime for the excited state of 8.3 seconds.}
\label{Fig:levels}
\end{figure}

Here we propose an optical clock based on a fine structure transition in Ba$^{4+}$. Figure~\ref{Fig:levels} presents the low-lying ($<\!50\,000$ cm$^{-1}$) energy spectrum of Ba$^{4+}$. These energy levels are experimentally known to $\sim\!1$ cm$^{-1}$ \cite{Rea83}. The lowest three electronic states are part of a $^3P_J$ fine structure manifold connected by optical transitions. Notably, this $^3P_J$ fine structure manifold has an anomalous energy ordering, being non-monotonic in $J$. Specifically, the ${^3P_1}$ state lies above, rather than between, the $^3P_0$ and $^3P_2$ states. As a consequence of this anomalous ordering, the first-excited state ($^3P_0$) cannot relax to the ground state ($^3P_2$) via magnetic dipole ($M1$) decay. It can relax, however, via electric quadrupole ($E2$) decay. As a general rule for optical transitions in atomic systems, allowed $E2$ decay is appreciably weaker than allowed $M1$ decay. This opens up the possibility of an optical clock based on the ${^3P_2}\rightarrow{^3P_0}$ fine structure transition, with the lifetime of the ${^3P_0}$ state being neither adversely short nor adversely long. At the same time, benefits of the fine structure transitions proposed in Ref.~\cite{YudTaiDer14} still apply, including theoretical tractability, low clock state multiplicity, and the potential for low systematic uncertainty.  In addition, the relatively low ionization energy ($\approx\!60$~eV) makes Ba$^{4+}$ an excellent candidate for production using compact ion sources \cite{MickKuhBuc18}. This is in contrast to many highly-charged ions identified as optical clock candidates in other works, which require ion sources operating at multi-keV electron beam energies (e.g., $\approx\!100$~keV for $^{207}$Pb$^{81+}$) \cite{NISTbasicASD}. Such ion sources tend to be located at dedicated high-energy facilities that are not well-suited for high-precision laser spectroscopy experiments.

Barium has several stable isotopes, all of which have nuclear spin $I=0$ or $I=3/2$. Both spin values offer distinct advantages. For $I=3/2$, the first order $E2$ shift can be suppressed, as described in Ref.~\cite{YudTaiDer14}. Namely, the $^3P_2$ hyperfine manifold contains an $F=1/2$ state, which has a vanishing $E2$ moment and may be taken as the lower clock state. Meanwhile, the $E2$ moment for the $^3P_0$ state ($F=I=3/2$) will be largely suppressed, as it arises from hyperfine mixing with $J\neq0$ electronic states~\cite{BelLeiIta17}. The $I=3/2$ isotopes are subject to a first order Zeeman (i.e., $M1$) shift, although this shift may be cancelled using averaging techniques~\cite{Ita00,DubMadBer05}. However, higher order effects involving $M1$ and $E2$ interactions are of concern, as these shifts arise primarily due to mixing within the $^3P_2$ hyperfine manifold. That is, $M1$ and $E2$ interactions couple the $^3P_2,\,F=1/2$ clock state to the nearby $^3P_2,\,F\neq1/2$ states, with small energy denominators entering the corresponding perturbation expressions. As a practical matter, we note that to mitigate line-pulling effects, a sufficiently large bias magnetic field would need to be applied to lift degeneracy of the $^3P_0,\,F=3/2$ clock state, which has a $g$-factor orders-of-magnitude smaller than the $^3P_2,\,F=1/2$ clock state. For the $I=0$ isotopes, the atomic structure is simpler, as there is no hyperfine structure. This implies lower clock state multiplicity (six substates, compared to 24). The first order Zeeman and $E2$ shifts may be appreciable, but they can be cancelled using averaging techniques~\cite{Ita00,DubMadBer05}. Importantly, with the absence of hyperfine structure, higher order shifts involving $M1$ and $E2$ interactions will be largely suppressed compared to the $I\neq0$ case. In the remainder, we assume the isotope $^{138}$Ba, which has $I=0$ and a natural abundance of 71.7\%~\cite{NISTbasicASD}.

To study this prospective clock in more detail, we perform {\it ab initio} relativistic many-body calculations. We treat Ba$^{4+}$ as a six-valence-electron system. The configuration interaction method is supplemented with many-body perturbation theory to account for correlations between the valence electrons as well as correlations with the core (CI+MBPT method). We start with a Dirac-Hartree-Fock (DHF) description of the core in absence of the valence electrons. The resulting DHF potential is used to generate single-electron orbitals, from which the many-electron basis states are constructed for computing the CI matrix. The basis is taken to include states with single and double excitations from a reference configuration (e.g., the $5s^25p^4$ ground state configuration). To restrict the basis and ensure a CI matrix of manageable size, high-energy states are treated perturbatively (the CIPT method). $E1$, $M1$, and $E2$ matrix elements are computed using the random-phase approximation (RPA). More details on these computational techniques can be found, e.g., in Refs.~\cite{DzuFlaKoz96,Dzu05,DzuBerHar17,DzuFlaSil87}.

In Figure~\ref{Fig:levels}, we present calculated decay rates for the lowest-lying states of Ba$^{4+}$, including all $M1$ and $E2$ decay channels. As expected, we observe that the $E2$ decay is generally weaker than the $M1$ decay. For the $^3P_0$ clock state, the calculated $E2$ decay rate is 0.12~s$^{-1}$, corresponding to a lifetime of 8.3 seconds. Using Eq.~(\ref{Eq:stabform}), we find a potential fractional instability of $\sigma_y(t)=4\times10^{-16}/\sqrt{t/\mathrm{s}}$. A fractional statistical uncertainty of $5\times10^{-19}$, for instance, could conceivably be obtained with just one week of averaging.

To first approximation, the blackbody radiation (BBR) frequency shift reads $-(1/2h)\left\langle E^2\right\rangle\Delta\alpha$, where $h$ is Planck's constant and $\left\langle E^2\right\rangle$ is the mean-squared electric field of the BBR. Here $\Delta\alpha=\alpha_e-\alpha_g$, where $\alpha_g$ and $\alpha_e$ are the static scalar polarizabilities of the ground and excited clock states, respectively. The mean-squared electric field satisfies $\left\langle E^2\right\rangle\approx \left(8.32~\text{V/cm}\right)^2\left(T/300~\text{K}\right)^4$, where $T$ is the temperature of the BBR environment. We compute $\alpha_g=4.4~{e^2a_B^2/E_H}$ and $\alpha_e=1.4~{e^2a_B^2/E_H}$, where $e$ is the elementary charge, $a_B$ is the Bohr radius, and $E_H$ is the Hartree unit of energy. The computational accuracy is estimated to be $\approx\!10\%$. At room temperature, the resulting $\Delta\alpha$ implies a fractional BBR shift of $7.6\times10^{-17}$. With its moderate ionization degree, room temperature operation could be feasible for Ba$^{4+}$. In this case, it's worth noting that experimental techniques are available to reduce uncertainty in $\Delta\alpha$~\cite{DubMadTib14,BarArnSaf19}. However, for a reduced ion loss rate (due to background gas collisions) and also to reduce BBR shift uncertainty, cryogenic operation would likely be desired (as demonstrated with Hg$^+$~\cite{RosHumSch08} and Ar$^{13+}$~\cite{MicLeoKin20}). In this case, the relatively small differential polarizability relaxes the need to precisely characterize the BBR environment. For example, constraining the temperature of the surroundings to $<\!55$ K would be sufficient to bound the fractional BBR shift to $<\!10^{-19}$. Dynamic and higher multipolar corrections~\cite{PorDer06} enter at the $10^{-21}$ level for room temperature operation and much less for cryogenic operation.

Second order Doppler (time-dilation) shifts due to ion motion are a dominant source of systematic uncertainty in ion clocks~\cite{BreCheHan19}. These time-dilation shifts are due to driven motion called excess micromotion (EMM), caused by trapping imperfections, and thermal (secular) motion due to the finite temperature of the ion in the trap. The magnitude and uncertainty of the secular motion shift can be reduced through sympathetic cooling with a logic ion~\cite{SchRosLan05}. To reduce the EMM-induced frequency shift and uncertainty, we propose operating the trap with an rf drive frequency tuned to the ``magic'' frequency, where the EMM-induced time-dilation shift is cancelled by the correlated scalar Stark shift. Such cancellation can only be realized for clock transitions possessing a negative differential polarizability, as is the case for Ba$^{4+}$. To lowest order, the magic rf frequency $\Omega_0/2\pi$ satisfies
$\Omega_{0}^{2}=-\left(h\nu_0/\Delta\alpha\right)\left(q_c/M_c c\right)^2$,
where $q_c$ and $M_c$ are the charge and mass of the clock ion and $c$ is the speed of light~\cite{BerMilBer98,DubMadZho13}. This technique has been successfully employed in singly-charged ions such as Ca$^+$~\cite{HuaGuaZen19} and Sr$^+$~\cite{DubMadTib14}, with Lu$^+$~\cite{ArnKaeRoy18} also being viable. For the case of $^{138}$Ba$^{4+}$, we find an experimentally-agreeable value $\Omega_{0}/2\pi=100$~MHz. We note that this rf frequency is orders of magnitude below all $E1$ transition frequencies from the clock states, justifying the use of the static values of the polarizabilities. The tensor Stark shift (not cancelled here) is addressed below.

With its relatively high atomic mass and moderate ionization degree, $^{138}$Ba$^{4+}$ has a charge-to-mass ratio comparable to $^{27}$Al$^+$ (only 22\% smaller). We propose using $^{40}$Ca$^{+}$ as a logic ion, which is currently being employed in the latest-generation Al$^+$ clocks at NIST and PTB~\cite{HanPelSch19}. It has a charge-to-mass ratio similar to $^{138}$Ba$^{4+}$ (14\% smaller), which helps suppress sympathetic laser-cooling inefficiencies~\cite{WanGebWol15}. The lasers required for laser cooling and readout operations are all diode-based, and $^{40}$Ca$^{+}$ has the added feature that it is possible to utilize electrically-induced transparency (EIT) cooling to cool several ions to near the motional ground state in a few hundred microseconds~\cite{LecMaiHem16}.

To estimate the time-dilation shift due to secular motion, we assume a linear Paul trap, similar to those constructed for the NIST Al$^+$ clocks \cite{BreCheHan19}, operating at the magic rf drive frequency.  For a $^{40}$Ca$^{+}$ ion with axial mode frequency $\approx\!1$~MHz and radial mode frequencies $\approx\!3.5$~MHz, the corresponding $^{138}$Ba$^{4+}$/$^{40}$Ca$^{+}$ mode frequencies range from $\approx\!1$~MHz to $\approx\!4$~MHz. We propose a ground-state-cooled (GSC) operation sequence where all modes are cooled to near the motional ground state prior to clock interrogation~\cite{CheBreCho17,BreCheHan19}.  For GSC performance similar to that achieved in the NIST Al$^{+}$ clock and modest improvements in motional heating rates, we estimate the secular motion time-dilation shift to be at the $10^{-18}$ fractional level for an interrogation time of 8.3~s, with a corresponding uncertainty at the low $10^{-19}$ level.

We assume a bias magnetic field $\mathbf{B}=B\hat{\mathbf{e}}_B$ lifts degeneracy in the $^3P_2$ clock state and defines the quantization axis. The ${^3P_2,m_J}\rightarrow{^3P_0}$ transition frequency reads
\begin{equation}
\nu\left(m_J\right)=\nu_0-m_Jf_{M1}-\left(m_J^2-2\right)f_{E2}+\delta\nu\left(m_J\right),
\label{Eq:M1E2}
\end{equation}
where the second and third terms on the right-hand-side account for the first order Zeeman shift ($\propto\!f_{M1}$) and the first order $E2$ shift ($\propto\!f_{E2}$). The last term, $\delta\nu\left(m_J\right)$, encapsulates all other frequency shifts. Here $f_{M1}=g\mu_BB/h$ and $f_{E2}=\Theta\varepsilon/4h$, where $g$ and $\Theta$ are the $g$-factor and $E2$ moment of the $^3P_2$ state, respectively, and $\mu_B$ is the Bohr magneton. We calculate $g=1.42$ and $\Theta=0.34\,ea_B^2$. The parameter $\varepsilon$ is given by $\varepsilon=\left(\hat{\mathbf{e}}_B\cdot\bm{\nabla}\right)^2\Phi$. Here $\Phi$ is the electrostatic potential due to the trap electrodes and the logic ion, and the right-hand-side is to be evaluated at the location of the clock ion~\cite{BelLeiIta17}. We note that $\varepsilon$ depends on the orientation of the magnetic field with respect to the trap frame.

Different strategies can be employed to eliminate the first order Zeeman and $E2$ shifts~\cite{Ita00,DubMadBer05}. One strategy is to use the $m_J=0$ substate and to average over three mutually-orthogonal magnetic field directions. The Zeeman shift vanishes for $m_J=0$, while the $E2$ shift vanishes due to the directional averaging (with the average value of $\varepsilon$ being $(1/3)\nabla^2\Phi=0$). In practice, however, some degree of non-orthogonality is inevitable, and constraining the uncancelled $E2$ shift may be challenging. An alternative strategy is to use a fixed magnetic field and to average over all $m_J$ values, which has the effect of cancelling both shifts. Using an appropriate weighting, the same outcome can be achieved with three transitions rather than all five. For example, introducing
\begin{equation}
\begin{gathered}
\nu_\mathrm{syn}=-\nu\left(0\right)+\nu\left(+1\right)+\nu\left(-1\right),
\\
\delta\nu_\mathrm{syn}=-\delta\nu\left(0\right)+\delta\nu\left(+1\right)+\delta\nu\left(-1\right),
\end{gathered}
\label{Eq:syn}
\end{equation}
it follows from Eq.~(\ref{Eq:M1E2}) that $\nu_\mathrm{syn}=\nu_0+\delta\nu_\mathrm{syn}$. That is, by synthesizing the frequency $\nu_\mathrm{syn}$, the first order Zeeman and $E2$ shifts are cancelled. We note that the $m_J$-dependencies appearing in Eq.~(\ref{Eq:M1E2}) are a consequence of the rank-1 (for $M1$) and rank-2 (for $E2$) tensor character of the shifts. It follows that shifts incorporated in $\delta\nu\left(m_J\right)$ that have rank-1 or rank-2 tensor character are likewise cancelled. An example includes the (rank-2) tensor Stark shift that arises due to trapping imperfections, with only the scalar Stark shift being cancelled by operation at the magic rf drive frequency.

Assuming the magnetic field to be nominally aligned with the trap axis, $\varepsilon$ satisfies $\varepsilon\approx2(\zeta_c/q_c)\mu_\pm\omega_\pm^2$, where $\omega_\pm/2\pi$ are the frequencies for the two modes of coupled axial motion, $\zeta_c=\left(3+q_c/q_l\right)/4$, $\zeta_l=\left(3+q_l/q_c\right)/4$, and
\begin{gather*}
\mu_\pm=\frac{M_cM_l}{\left(M_c\zeta_l+M_l\zeta_c\right)\pm\sqrt{\left(M_c\zeta_l-M_l\zeta_c\right)^2+M_cM_l}}.
\end{gather*}
Here $q_l$ and $M_l$ are the charge and mass of the logic ion. The axial mode frequencies satisfy $\omega_+/\omega_-=\sqrt{\mu_-/\mu_+}$, with the factor $\mu_\pm\omega_\pm^2$ in the expression for $\varepsilon$ being independent of the mode. For the $^{138}$Ba$^{4+}$/$^{40}$Ca$^+$ ion pair, we have $\omega_+/\omega_-=1.66$. Assuming $\omega_-/2\pi\approx1$~MHz and using our calculated value of $\Theta$, we find $f_{E2}/\nu_0\approx3.5\times10^{-15}$. While other magnetic field orientations are acceptable (with $f_{E2}$ being of the same order or less), there is a practical advantage to aligning the magnetic field along the trap axis, with $f_{E2}$ having suppressed sensitivity to directional variations in the field.

Equation (\ref{Eq:M1E2}) implicitly assumes that the $M1$ interaction dominates over the $E2$ interaction, with $\mathbf{B}$ defining the quantization axis and $m_J$ being treated as a ``good'' quantum number. However, the fact that the $E2$ interaction is generally non-diagonal in $m_J$ must be considered. Corresponding corrections can be incorporated into $\delta\nu\left(m_J\right)$ of Eq.~(\ref{Eq:M1E2}). The leading corrections are of order $f_{E2}^2/f_{M1}$ (or less, depending on the radial symmetry of the trap). Assuming $B\approx5~\mu$T, we have $\left|f_{E2}^2/f_{M1}\nu_0\right|\approx10^{-20}$. These corrections are necessarily odd with respect to $m_J$ and, from Eq.~(\ref{Eq:syn}), cancel out of $\delta\nu_\mathrm{syn}$. The next-leading corrections are of order $f_{E2}^3/f_{M1}^2$ (or less) and will be negligible.

Aside from the Zeeman and $E2$ shifts considered thus far, there are also higher order shifts involving the $M1$ and $E2$ interactions. Of these, only the second order Zeeman shift is potentially non-negligible. The second order Zeeman shift can be partitioned into rank-0 and rank-2 tensor parts, with each contributing comparably to $\delta\nu\left(m_J\right)$. The rank-2 part cancels out of $\delta\nu_\mathrm{syn}$, as noted above. Meanwhile, the contribution to $\delta\nu_\mathrm{syn}$ from the scalar (rank-0) part can be written $\beta f_{M1}^2$, where we calculate the factor $\beta$ to be $\beta=-6.8\times10^{-16}~\text{Hz}^{-1}$. For $B\approx5~\mu$T, we find $\beta f_{M1}^2/\nu_0\approx-2\times10^{-20}$. We note that in synthesizing $\nu_\mathrm{syn}$, the quantities $f_{M1}$ and $f_{E2}$ are essentially obtained at no cost. In particular, the relationships $f_{M1}\approx-(1/2)\left[\nu\left(+1\right)-\nu\left(-1\right)\right]$ and $f_{E2}\approx\nu\left(0\right)-(1/2)\left[\nu\left(+1\right)+\nu\left(-1\right)\right]$ follow from Eq.~(\ref{Eq:M1E2}). This permits a real-time assessment of the second order Zeeman shift, if only to ensure that it remains negligibly small, as well as an assessment of the stability of the first order Zeeman and $E2$ shifts.

If the ion can be initialized in the desired $^3P_2,m_J$ substate with high fidelity using quantum logic techniques, then line-pulling is inconsequential due to the non-degeneracy of the $^3P_0$ state. However, infidelity in the state initialization could open the door to line-pulling. For $B\approx 5~\mu$T, the transition frequencies $\nu\left(m_J\right)$ with different $m_J$ are separated by $\approx\!100$ kHz. Assuming Ramsey pulse durations of $\gtrsim10$ ms, line-pulling is constrained to $\lesssim\!10^{-22}$, fractionally.

Given the calculated $E2$ transition matrix element between the clock states (which can be inferred from the $E2$ decay rate) and the calculated differential polarizability between the clock states, we estimate that driving a $\pi/2$-pulse in $\gtrsim\!10$~ms would result in a fractional Stark shift of $\lesssim\!10^{-20}$ during the pulse. Consequently, exotic probing schemes~\cite{YudTaiOat10,SanHunLan18} would be unnecessary.

Fine structure energy splittings nominally scale as $\alpha^2E_H$. This implies that $K$, the factor that quantifies sensitivity of the clock transition to $\alpha$-variation~\cite{SafBudDeM18}, nominally equals two for fine structure transitions. However, the anomalous ordering of the fine structure manifold suggests that relativistic mixing between states plays a prominent role in the present case. We calculate $K=1.04$, indicating that this mixing reduces $K$ to roughly half the nominal value. For comparison, $K=0.008$ for Al$^{+}$ and $K=-5.95$ for Yb$^{+}$ (electric octupole transition), with all other demonstrated optical clocks having an intermediate value of $\left|K\right|$~\cite{SafBudDeM18}. Two clocks are required for observing $\alpha$-variation, with the sensitivity factor for the dimensionless frequency ratio equal to the difference in $K$ values. Noting the opposite sign of $K$ for Ba$^{4+}$ and Yb$^{+}$, these clocks would allow for a sensitive probe of $\alpha$-variation. We further note that large relativistic effects imply high sensitivity to violation of Einstein's equivalence principle and local Lorentz invariance~\cite{DzuFlaSch18}. All these effects (including $\alpha$-variation) might be manifestations of the interaction of atomic electrons with low-mass scalar dark matter~\cite{DerPos14,ArvHuaVan15,VanLeeBou15,StaFla15}.

In conclusion, we propose an optical clock based on the ${^3P_2}\rightarrow{^3P_0}$ fine structure transition in Ba$^{4+}$. This transition shares the benefits of the fine structure transitions considered in Ref.~\cite{YudTaiDer14}, including theoretical tractability, low clock state-multiplicity, and potential for low systematic uncertainty. In contrast to the transitions considered in Ref.~\cite{YudTaiDer14}, the excited clock state lacks an $M1$ decay channel. Instead the excited clock state relaxes by $E2$ decay with a lifetime of several seconds. This lifetime accommodates low statistical uncertainty for practical averaging times. Furthermore, Ba$^{4+}$ is an excellent candidate for production in a low-energy ion source and possesses an experimentally-convenient magic rf trapping frequency, which can be exploited to cancel Stark and excess micromotion shifts. This magic rf trapping frequency could also help enable multi-ion clock operation. Finally, we note that isoelectronic systems with ionization degree up to six may also be of interest, as they have similar electronic structure to Ba$^{4+}$. Of these, Xe$^{2+}$ and Ce$^{6+}$ possess nuclear spin-zero isotopes.

\begin{acknowledgments}
The authors thank May Kim and Yuri Ralchenko for their careful reading of the manuscript. This work was supported by NIST
Physical Measurement Laboratory (KB), the Australian Research Council (VAD), and Colorado State University (SMB).
\end{acknowledgments}


\begin{thebibliography}{78}%
\makeatletter
\providecommand \@ifxundefined [1]{%
 \@ifx{#1\undefined}
}%
\providecommand \@ifnum [1]{%
 \ifnum #1\expandafter \@firstoftwo
 \else \expandafter \@secondoftwo
 \fi
}%
\providecommand \@ifx [1]{%
 \ifx #1\expandafter \@firstoftwo
 \else \expandafter \@secondoftwo
 \fi
}%
\providecommand \natexlab [1]{#1}%
\providecommand \enquote  [1]{``#1''}%
\providecommand \bibnamefont  [1]{#1}%
\providecommand \bibfnamefont [1]{#1}%
\providecommand \citenamefont [1]{#1}%
\providecommand \href@noop [0]{\@secondoftwo}%
\providecommand \href [0]{\begingroup \@sanitize@url \@href}%
\providecommand \@href[1]{\@@startlink{#1}\@@href}%
\providecommand \@@href[1]{\endgroup#1\@@endlink}%
\providecommand \@sanitize@url [0]{\catcode `\\12\catcode `\$12\catcode
  `\&12\catcode `\#12\catcode `\^12\catcode `\_12\catcode `\%12\relax}%
\providecommand \@@startlink[1]{}%
\providecommand \@@endlink[0]{}%
\providecommand \url  [0]{\begingroup\@sanitize@url \@url }%
\providecommand \@url [1]{\endgroup\@href {#1}{\urlprefix }}%
\providecommand \urlprefix  [0]{URL }%
\providecommand \Eprint [0]{\href }%
\providecommand \doibase [0]{http://dx.doi.org/}%
\providecommand \selectlanguage [0]{\@gobble}%
\providecommand \bibinfo  [0]{\@secondoftwo}%
\providecommand \bibfield  [0]{\@secondoftwo}%
\providecommand \translation [1]{[#1]}%
\providecommand \BibitemOpen [0]{}%
\providecommand \bibitemStop [0]{}%
\providecommand \bibitemNoStop [0]{.\EOS\space}%
\providecommand \EOS [0]{\spacefactor3000\relax}%
\providecommand \BibitemShut  [1]{\csname bibitem#1\endcsname}%
\let\auto@bib@innerbib\@empty
\bibitem [{\citenamefont {Ludlow}\ \emph {et~al.}(2015)\citenamefont {Ludlow},
  \citenamefont {Boyd}, \citenamefont {Ye}, \citenamefont {Peik},\ and\
  \citenamefont {Schmidt}}]{LudBoyYe15}%
  \BibitemOpen
  \bibfield  {author} {\bibinfo {author} {\bibfnamefont {A.~D.}\ \bibnamefont
  {Ludlow}}, \bibinfo {author} {\bibfnamefont {M.~M.}\ \bibnamefont {Boyd}},
  \bibinfo {author} {\bibfnamefont {J.}~\bibnamefont {Ye}}, \bibinfo {author}
  {\bibfnamefont {E.}~\bibnamefont {Peik}}, \ and\ \bibinfo {author}
  {\bibfnamefont {P.~O.}\ \bibnamefont {Schmidt}},\ }\href {\doibase
  10.1103/RevModPhys.87.637} {\bibfield  {journal} {\bibinfo  {journal} {Rev.
  Mod. Phys.}\ }\textbf {\bibinfo {volume} {87}},\ \bibinfo {pages} {637}
  (\bibinfo {year} {2015})}\BibitemShut {NoStop}%
\bibitem [{\citenamefont {Ushijima}\ \emph {et~al.}(2015)\citenamefont
  {Ushijima}, \citenamefont {Takamoto}, \citenamefont {Das}, \citenamefont
  {Ohkubo},\ and\ \citenamefont {Katori}}]{UshTakDas15}%
  \BibitemOpen
  \bibfield  {author} {\bibinfo {author} {\bibfnamefont {I.}~\bibnamefont
  {Ushijima}}, \bibinfo {author} {\bibfnamefont {M.}~\bibnamefont {Takamoto}},
  \bibinfo {author} {\bibfnamefont {M.}~\bibnamefont {Das}}, \bibinfo {author}
  {\bibfnamefont {T.}~\bibnamefont {Ohkubo}}, \ and\ \bibinfo {author}
  {\bibfnamefont {H.}~\bibnamefont {Katori}},\ }\href {\doibase
  10.1038/nphoton.2015.5} {\bibfield  {journal} {\bibinfo  {journal} {Nat.
  Photon.}\ }\textbf {\bibinfo {volume} {9}},\ \bibinfo {pages} {185} (\bibinfo
  {year} {2015})}\BibitemShut {NoStop}%
\bibitem [{\citenamefont {Nemitz}\ \emph {et~al.}(2016)\citenamefont {Nemitz},
  \citenamefont {Ohkubo}, \citenamefont {Takamoto}, \citenamefont {Ushijima},
  \citenamefont {Das}, \citenamefont {Ohmae},\ and\ \citenamefont
  {Katori}}]{NemOhkTak16}%
  \BibitemOpen
  \bibfield  {author} {\bibinfo {author} {\bibfnamefont {N.}~\bibnamefont
  {Nemitz}}, \bibinfo {author} {\bibfnamefont {T.}~\bibnamefont {Ohkubo}},
  \bibinfo {author} {\bibfnamefont {M.}~\bibnamefont {Takamoto}}, \bibinfo
  {author} {\bibfnamefont {I.}~\bibnamefont {Ushijima}}, \bibinfo {author}
  {\bibfnamefont {M.}~\bibnamefont {Das}}, \bibinfo {author} {\bibfnamefont
  {N.}~\bibnamefont {Ohmae}}, \ and\ \bibinfo {author} {\bibfnamefont
  {H.}~\bibnamefont {Katori}},\ }\href {\doibase 10.1038/nphoton.2016.20}
  {\bibfield  {journal} {\bibinfo  {journal} {Nat. Photon.}\ }\textbf {\bibinfo
  {volume} {10}},\ \bibinfo {pages} {258} (\bibinfo {year} {2016})}\BibitemShut
  {NoStop}%
\bibitem [{\citenamefont {McGrew}\ \emph {et~al.}(2018)\citenamefont {McGrew},
  \citenamefont {Zhang}, \citenamefont {Fasano}, \citenamefont {Sch\"{a}ffer},
  \citenamefont {Beloy}, \citenamefont {Nicolodi}, \citenamefont {Brown},
  \citenamefont {Hinkley}, \citenamefont {Milani}, \citenamefont {Schioppo},
  \citenamefont {Yoon},\ and\ \citenamefont {Ludlow}}]{McGZhaFas18}%
  \BibitemOpen
  \bibfield  {author} {\bibinfo {author} {\bibfnamefont {W.~F.}\ \bibnamefont
  {McGrew}}, \bibinfo {author} {\bibfnamefont {X.}~\bibnamefont {Zhang}},
  \bibinfo {author} {\bibfnamefont {R.~J.}\ \bibnamefont {Fasano}}, \bibinfo
  {author} {\bibfnamefont {S.~A.}\ \bibnamefont {Sch\"{a}ffer}}, \bibinfo
  {author} {\bibfnamefont {K.}~\bibnamefont {Beloy}}, \bibinfo {author}
  {\bibfnamefont {D.}~\bibnamefont {Nicolodi}}, \bibinfo {author}
  {\bibfnamefont {R.~C.}\ \bibnamefont {Brown}}, \bibinfo {author}
  {\bibfnamefont {N.}~\bibnamefont {Hinkley}}, \bibinfo {author} {\bibfnamefont
  {G.}~\bibnamefont {Milani}}, \bibinfo {author} {\bibfnamefont
  {M.}~\bibnamefont {Schioppo}}, \bibinfo {author} {\bibfnamefont {T.~H.}\
  \bibnamefont {Yoon}}, \ and\ \bibinfo {author} {\bibfnamefont {A.~D.}\
  \bibnamefont {Ludlow}},\ }\href {\doibase 10.1038/s41586-018-0738-2}
  {\bibfield  {journal} {\bibinfo  {journal} {Nature}\ }\textbf {\bibinfo
  {volume} {564}},\ \bibinfo {pages} {87} (\bibinfo {year} {2018})}\BibitemShut
  {NoStop}%
\bibitem [{\citenamefont {Bothwell}\ \emph {et~al.}(2019)\citenamefont
  {Bothwell}, \citenamefont {Kedar}, \citenamefont {Oelker}, \citenamefont
  {Robinson}, \citenamefont {Bromley}, \citenamefont {Tew}, \citenamefont
  {Ye},\ and\ \citenamefont {Kennedy}}]{BotKedOel19}%
  \BibitemOpen
  \bibfield  {author} {\bibinfo {author} {\bibfnamefont {T.}~\bibnamefont
  {Bothwell}}, \bibinfo {author} {\bibfnamefont {D.}~\bibnamefont {Kedar}},
  \bibinfo {author} {\bibfnamefont {E.}~\bibnamefont {Oelker}}, \bibinfo
  {author} {\bibfnamefont {J.~M.}\ \bibnamefont {Robinson}}, \bibinfo {author}
  {\bibfnamefont {S.~L.}\ \bibnamefont {Bromley}}, \bibinfo {author}
  {\bibfnamefont {W.~L.}\ \bibnamefont {Tew}}, \bibinfo {author} {\bibfnamefont
  {J.}~\bibnamefont {Ye}}, \ and\ \bibinfo {author} {\bibfnamefont {C.~J.}\
  \bibnamefont {Kennedy}},\ }\href {\doibase 10.1088/1681-7575/ab4089}
  {\bibfield  {journal} {\bibinfo  {journal} {Metrologia}\ }\textbf {\bibinfo
  {volume} {56}},\ \bibinfo {pages} {065004} (\bibinfo {year}
  {2019})}\BibitemShut {NoStop}%
\bibitem [{\citenamefont {Pizzocaro}\ \emph {et~al.}()\citenamefont
  {Pizzocaro}, \citenamefont {Bregolin}, \citenamefont {Barbieri},
  \citenamefont {Rauf}, \citenamefont {Levi},\ and\ \citenamefont
  {Calonico}}]{PizBreBarinpress}%
  \BibitemOpen
  \bibfield  {author} {\bibinfo {author} {\bibfnamefont {M.}~\bibnamefont
  {Pizzocaro}}, \bibinfo {author} {\bibfnamefont {F.}~\bibnamefont {Bregolin}},
  \bibinfo {author} {\bibfnamefont {P.}~\bibnamefont {Barbieri}}, \bibinfo
  {author} {\bibfnamefont {B.}~\bibnamefont {Rauf}}, \bibinfo {author}
  {\bibfnamefont {F.}~\bibnamefont {Levi}}, \ and\ \bibinfo {author}
  {\bibfnamefont {D.}~\bibnamefont {Calonico}},\ }\href@noop {} {\bibfield
  {journal} {\bibinfo  {journal} {Metrologia}\ }}\bibinfo {note} {{(in
  press)}}\BibitemShut {NoStop}%
\bibitem [{\citenamefont {Madej}\ \emph {et~al.}(2012)\citenamefont {Madej},
  \citenamefont {Dub\'e}, \citenamefont {Zhou}, \citenamefont {Bernard},\ and\
  \citenamefont {Gertsvolf}}]{MadDubZho12}%
  \BibitemOpen
  \bibfield  {author} {\bibinfo {author} {\bibfnamefont {A.~A.}\ \bibnamefont
  {Madej}}, \bibinfo {author} {\bibfnamefont {P.}~\bibnamefont {Dub\'e}},
  \bibinfo {author} {\bibfnamefont {Z.}~\bibnamefont {Zhou}}, \bibinfo {author}
  {\bibfnamefont {J.~E.}\ \bibnamefont {Bernard}}, \ and\ \bibinfo {author}
  {\bibfnamefont {M.}~\bibnamefont {Gertsvolf}},\ }\href {\doibase
  10.1103/PhysRevLett.109.203002} {\bibfield  {journal} {\bibinfo  {journal}
  {Phys. Rev. Lett.}\ }\textbf {\bibinfo {volume} {109}},\ \bibinfo {pages}
  {203002} (\bibinfo {year} {2012})}\BibitemShut {NoStop}%
\bibitem [{\citenamefont {Huntemann}\ \emph {et~al.}(2016)\citenamefont
  {Huntemann}, \citenamefont {Sanner}, \citenamefont {Lipphardt}, \citenamefont
  {Tamm},\ and\ \citenamefont {Peik}}]{HunSanLip16}%
  \BibitemOpen
  \bibfield  {author} {\bibinfo {author} {\bibfnamefont {N.}~\bibnamefont
  {Huntemann}}, \bibinfo {author} {\bibfnamefont {C.}~\bibnamefont {Sanner}},
  \bibinfo {author} {\bibfnamefont {B.}~\bibnamefont {Lipphardt}}, \bibinfo
  {author} {\bibfnamefont {C.}~\bibnamefont {Tamm}}, \ and\ \bibinfo {author}
  {\bibfnamefont {E.}~\bibnamefont {Peik}},\ }\href {\doibase
  10.1103/PhysRevLett.116.063001} {\bibfield  {journal} {\bibinfo  {journal}
  {Phys. Rev. Lett.}\ }\textbf {\bibinfo {volume} {116}},\ \bibinfo {pages}
  {063001} (\bibinfo {year} {2016})}\BibitemShut {NoStop}%
\bibitem [{\citenamefont {Huang}\ \emph {et~al.}(2017)\citenamefont {Huang},
  \citenamefont {Guan}, \citenamefont {Bian}, \citenamefont {Ma}, \citenamefont
  {Liang}, \citenamefont {Li},\ and\ \citenamefont {Gao}}]{HuaGuaBia17}%
  \BibitemOpen
  \bibfield  {author} {\bibinfo {author} {\bibfnamefont {Y.}~\bibnamefont
  {Huang}}, \bibinfo {author} {\bibfnamefont {H.}~\bibnamefont {Guan}},
  \bibinfo {author} {\bibfnamefont {W.}~\bibnamefont {Bian}}, \bibinfo {author}
  {\bibfnamefont {L.}~\bibnamefont {Ma}}, \bibinfo {author} {\bibfnamefont
  {K.}~\bibnamefont {Liang}}, \bibinfo {author} {\bibfnamefont
  {T.}~\bibnamefont {Li}}, \ and\ \bibinfo {author} {\bibfnamefont
  {K.}~\bibnamefont {Gao}},\ }\href {\doibase 10.1007/s00340-017-6731-x}
  {\bibfield  {journal} {\bibinfo  {journal} {Appl. Phys. B}\ }\textbf
  {\bibinfo {volume} {123}},\ \bibinfo {pages} {166} (\bibinfo {year}
  {2017})}\BibitemShut {NoStop}%
\bibitem [{\citenamefont {Baynham}\ \emph {et~al.}(2018)\citenamefont
  {Baynham}, \citenamefont {Godun}, \citenamefont {Jones}, \citenamefont
  {King}, \citenamefont {Nisbet-Jones}, \citenamefont {Baynes}, \citenamefont
  {Rolland}, \citenamefont {Baird}, \citenamefont {Bongs}, \citenamefont
  {Gill},\ and\ \citenamefont {Margolis}}]{BayGodJon18}%
  \BibitemOpen
  \bibfield  {author} {\bibinfo {author} {\bibfnamefont {C.~F.~A.}\
  \bibnamefont {Baynham}}, \bibinfo {author} {\bibfnamefont {R.~M.}\
  \bibnamefont {Godun}}, \bibinfo {author} {\bibfnamefont {J.~M.}\ \bibnamefont
  {Jones}}, \bibinfo {author} {\bibfnamefont {S.~A.}\ \bibnamefont {King}},
  \bibinfo {author} {\bibfnamefont {P.~B.~R.}\ \bibnamefont {Nisbet-Jones}},
  \bibinfo {author} {\bibfnamefont {F.}~\bibnamefont {Baynes}}, \bibinfo
  {author} {\bibfnamefont {A.}~\bibnamefont {Rolland}}, \bibinfo {author}
  {\bibfnamefont {P.~E.~G.}\ \bibnamefont {Baird}}, \bibinfo {author}
  {\bibfnamefont {K.}~\bibnamefont {Bongs}}, \bibinfo {author} {\bibfnamefont
  {P.}~\bibnamefont {Gill}}, \ and\ \bibinfo {author} {\bibfnamefont {H.~S.}\
  \bibnamefont {Margolis}},\ }\href {\doibase 10.1080/09500340.2017.1384514}
  {\bibfield  {journal} {\bibinfo  {journal} {J. Mod. Optics}\ }\textbf
  {\bibinfo {volume} {65}},\ \bibinfo {pages} {585} (\bibinfo {year}
  {2018})}\BibitemShut {NoStop}%
\bibitem [{\citenamefont {Brewer}\ \emph {et~al.}(2019)\citenamefont {Brewer},
  \citenamefont {Chen}, \citenamefont {Hankin}, \citenamefont {Clements},
  \citenamefont {Chou}, \citenamefont {Wineland}, \citenamefont {Hume},\ and\
  \citenamefont {Leibrandt}}]{BreCheHan19}%
  \BibitemOpen
  \bibfield  {author} {\bibinfo {author} {\bibfnamefont {S.~M.}\ \bibnamefont
  {Brewer}}, \bibinfo {author} {\bibfnamefont {J.-S.}\ \bibnamefont {Chen}},
  \bibinfo {author} {\bibfnamefont {A.~M.}\ \bibnamefont {Hankin}}, \bibinfo
  {author} {\bibfnamefont {E.~R.}\ \bibnamefont {Clements}}, \bibinfo {author}
  {\bibfnamefont {C.~W.}\ \bibnamefont {Chou}}, \bibinfo {author}
  {\bibfnamefont {D.~J.}\ \bibnamefont {Wineland}}, \bibinfo {author}
  {\bibfnamefont {D.~B.}\ \bibnamefont {Hume}}, \ and\ \bibinfo {author}
  {\bibfnamefont {D.~R.}\ \bibnamefont {Leibrandt}},\ }\href {\doibase
  10.1103/PhysRevLett.123.033201} {\bibfield  {journal} {\bibinfo  {journal}
  {Phys. Rev. Lett.}\ }\textbf {\bibinfo {volume} {123}},\ \bibinfo {pages}
  {033201} (\bibinfo {year} {2019})}\BibitemShut {NoStop}%
\bibitem [{\citenamefont {Schmidt}\ \emph {et~al.}(2005)\citenamefont
  {Schmidt}, \citenamefont {Rosenband}, \citenamefont {Langer}, \citenamefont
  {Itano}, \citenamefont {Bergquist},\ and\ \citenamefont
  {Wineland}}]{SchRosLan05}%
  \BibitemOpen
  \bibfield  {author} {\bibinfo {author} {\bibfnamefont {P.~O.}\ \bibnamefont
  {Schmidt}}, \bibinfo {author} {\bibfnamefont {T.}~\bibnamefont {Rosenband}},
  \bibinfo {author} {\bibfnamefont {C.}~\bibnamefont {Langer}}, \bibinfo
  {author} {\bibfnamefont {W.~M.}\ \bibnamefont {Itano}}, \bibinfo {author}
  {\bibfnamefont {J.~C.}\ \bibnamefont {Bergquist}}, \ and\ \bibinfo {author}
  {\bibfnamefont {D.~J.}\ \bibnamefont {Wineland}},\ }\href {\doibase
  10.1126/science.1114375} {\bibfield  {journal} {\bibinfo  {journal}
  {Science}\ }\textbf {\bibinfo {volume} {309}},\ \bibinfo {pages} {749}
  (\bibinfo {year} {2005})}\BibitemShut {NoStop}%
\bibitem [{\citenamefont {Safronova}\ \emph {et~al.}(2018)\citenamefont
  {Safronova}, \citenamefont {Budker}, \citenamefont {DeMille}, \citenamefont
  {Kimball}, \citenamefont {Derevianko},\ and\ \citenamefont
  {Clark}}]{SafBudDeM18}%
  \BibitemOpen
  \bibfield  {author} {\bibinfo {author} {\bibfnamefont {M.~S.}\ \bibnamefont
  {Safronova}}, \bibinfo {author} {\bibfnamefont {D.}~\bibnamefont {Budker}},
  \bibinfo {author} {\bibfnamefont {D.}~\bibnamefont {DeMille}}, \bibinfo
  {author} {\bibfnamefont {D.~F.~J.}\ \bibnamefont {Kimball}}, \bibinfo
  {author} {\bibfnamefont {A.}~\bibnamefont {Derevianko}}, \ and\ \bibinfo
  {author} {\bibfnamefont {C.~W.}\ \bibnamefont {Clark}},\ }\href {\doibase
  10.1103/RevModPhys.90.025008} {\bibfield  {journal} {\bibinfo  {journal}
  {Rev. Mod. Phys.}\ }\textbf {\bibinfo {volume} {90}},\ \bibinfo {pages}
  {025008} (\bibinfo {year} {2018})}\BibitemShut {NoStop}%
\bibitem [{\citenamefont {Mehlst\"aubler}\ \emph {et~al.}(2018)\citenamefont
  {Mehlst\"aubler}, \citenamefont {Grosche}, \citenamefont {Lisdat},
  \citenamefont {Schmidt},\ and\ \citenamefont {Denker}}]{MehGroLis18}%
  \BibitemOpen
  \bibfield  {author} {\bibinfo {author} {\bibfnamefont {T.~E.}\ \bibnamefont
  {Mehlst\"aubler}}, \bibinfo {author} {\bibfnamefont {G.}~\bibnamefont
  {Grosche}}, \bibinfo {author} {\bibfnamefont {C.}~\bibnamefont {Lisdat}},
  \bibinfo {author} {\bibfnamefont {P.~O.}\ \bibnamefont {Schmidt}}, \ and\
  \bibinfo {author} {\bibfnamefont {H.}~\bibnamefont {Denker}},\ }\href
  {\doibase 10.1088/1361-6633/aab409} {\bibfield  {journal} {\bibinfo
  {journal} {Rep. Progress Phys.}\ }\textbf {\bibinfo {volume} {81}},\ \bibinfo
  {pages} {064401} (\bibinfo {year} {2018})}\BibitemShut {NoStop}%
\bibitem [{\citenamefont {Schiller}(2007)}]{Sch07}%
  \BibitemOpen
  \bibfield  {author} {\bibinfo {author} {\bibfnamefont {S.}~\bibnamefont
  {Schiller}},\ }\href {\doibase 10.1103/PhysRevLett.98.180801} {\bibfield
  {journal} {\bibinfo  {journal} {Phys. Rev. Lett.}\ }\textbf {\bibinfo
  {volume} {98}},\ \bibinfo {pages} {180801} (\bibinfo {year}
  {2007})}\BibitemShut {NoStop}%
\bibitem [{\citenamefont {Berengut}\ \emph {et~al.}(2010)\citenamefont
  {Berengut}, \citenamefont {Dzuba},\ and\ \citenamefont
  {Flambaum}}]{BerDzuFla10}%
  \BibitemOpen
  \bibfield  {author} {\bibinfo {author} {\bibfnamefont {J.~C.}\ \bibnamefont
  {Berengut}}, \bibinfo {author} {\bibfnamefont {V.~A.}\ \bibnamefont {Dzuba}},
  \ and\ \bibinfo {author} {\bibfnamefont {V.~V.}\ \bibnamefont {Flambaum}},\
  }\href {\doibase 10.1103/PhysRevLett.105.120801} {\bibfield  {journal}
  {\bibinfo  {journal} {Phys. Rev. Lett.}\ }\textbf {\bibinfo {volume} {105}},\
  \bibinfo {pages} {120801} (\bibinfo {year} {2010})}\BibitemShut {NoStop}%
\bibitem [{\citenamefont {Berengut}\ \emph
  {et~al.}(2011{\natexlab{a}})\citenamefont {Berengut}, \citenamefont {Dzuba},\
  and\ \citenamefont {Flambaum}}]{BerDzuFla11a}%
  \BibitemOpen
  \bibfield  {author} {\bibinfo {author} {\bibfnamefont {J.~C.}\ \bibnamefont
  {Berengut}}, \bibinfo {author} {\bibfnamefont {V.~A.}\ \bibnamefont {Dzuba}},
  \ and\ \bibinfo {author} {\bibfnamefont {V.~V.}\ \bibnamefont {Flambaum}},\
  }\href {\doibase 10.1103/PhysRevA.84.054501} {\bibfield  {journal} {\bibinfo
  {journal} {Phys. Rev. A}\ }\textbf {\bibinfo {volume} {84}},\ \bibinfo
  {pages} {054501} (\bibinfo {year} {2011}{\natexlab{a}})}\BibitemShut
  {NoStop}%
\bibitem [{\citenamefont {Berengut}\ \emph
  {et~al.}(2011{\natexlab{b}})\citenamefont {Berengut}, \citenamefont {Dzuba},
  \citenamefont {Flambaum},\ and\ \citenamefont {Ong}}]{BerDzuFla11b}%
  \BibitemOpen
  \bibfield  {author} {\bibinfo {author} {\bibfnamefont {J.~C.}\ \bibnamefont
  {Berengut}}, \bibinfo {author} {\bibfnamefont {V.~A.}\ \bibnamefont {Dzuba}},
  \bibinfo {author} {\bibfnamefont {V.~V.}\ \bibnamefont {Flambaum}}, \ and\
  \bibinfo {author} {\bibfnamefont {A.}~\bibnamefont {Ong}},\ }\href {\doibase
  10.1103/PhysRevLett.106.210802} {\bibfield  {journal} {\bibinfo  {journal}
  {Phys. Rev. Lett.}\ }\textbf {\bibinfo {volume} {106}},\ \bibinfo {pages}
  {210802} (\bibinfo {year} {2011}{\natexlab{b}})}\BibitemShut {NoStop}%
\bibitem [{\citenamefont {Berengut}\ \emph
  {et~al.}(2012{\natexlab{a}})\citenamefont {Berengut}, \citenamefont {Dzuba},
  \citenamefont {Flambaum},\ and\ \citenamefont {Ong}}]{BerDzuFla12a}%
  \BibitemOpen
  \bibfield  {author} {\bibinfo {author} {\bibfnamefont {J.~C.}\ \bibnamefont
  {Berengut}}, \bibinfo {author} {\bibfnamefont {V.~A.}\ \bibnamefont {Dzuba}},
  \bibinfo {author} {\bibfnamefont {V.~V.}\ \bibnamefont {Flambaum}}, \ and\
  \bibinfo {author} {\bibfnamefont {A.}~\bibnamefont {Ong}},\ }\href {\doibase
  10.1103/PhysRevA.86.022517} {\bibfield  {journal} {\bibinfo  {journal} {Phys.
  Rev. A}\ }\textbf {\bibinfo {volume} {86}},\ \bibinfo {pages} {022517}
  (\bibinfo {year} {2012}{\natexlab{a}})}\BibitemShut {NoStop}%
\bibitem [{\citenamefont {Berengut}\ \emph
  {et~al.}(2012{\natexlab{b}})\citenamefont {Berengut}, \citenamefont {Dzuba},
  \citenamefont {Flambaum},\ and\ \citenamefont {Ong}}]{BerDzuFla12b}%
  \BibitemOpen
  \bibfield  {author} {\bibinfo {author} {\bibfnamefont {J.~C.}\ \bibnamefont
  {Berengut}}, \bibinfo {author} {\bibfnamefont {V.~A.}\ \bibnamefont {Dzuba}},
  \bibinfo {author} {\bibfnamefont {V.~V.}\ \bibnamefont {Flambaum}}, \ and\
  \bibinfo {author} {\bibfnamefont {A.}~\bibnamefont {Ong}},\ }\href {\doibase
  10.1103/PhysRevLett.109.070802} {\bibfield  {journal} {\bibinfo  {journal}
  {Phys. Rev. Lett.}\ }\textbf {\bibinfo {volume} {109}},\ \bibinfo {pages}
  {070802} (\bibinfo {year} {2012}{\natexlab{b}})}\BibitemShut {NoStop}%
\bibitem [{\citenamefont {Derevianko}\ \emph {et~al.}(2012)\citenamefont
  {Derevianko}, \citenamefont {Dzuba},\ and\ \citenamefont
  {Flambaum}}]{DerDzuFla12}%
  \BibitemOpen
  \bibfield  {author} {\bibinfo {author} {\bibfnamefont {A.}~\bibnamefont
  {Derevianko}}, \bibinfo {author} {\bibfnamefont {V.~A.}\ \bibnamefont
  {Dzuba}}, \ and\ \bibinfo {author} {\bibfnamefont {V.~V.}\ \bibnamefont
  {Flambaum}},\ }\href {\doibase 10.1103/PhysRevLett.109.180801} {\bibfield
  {journal} {\bibinfo  {journal} {Phys. Rev. Lett.}\ }\textbf {\bibinfo
  {volume} {109}},\ \bibinfo {pages} {180801} (\bibinfo {year}
  {2012})}\BibitemShut {NoStop}%
\bibitem [{\citenamefont {Dzuba}\ \emph
  {et~al.}(2012{\natexlab{a}})\citenamefont {Dzuba}, \citenamefont
  {Derevianko},\ and\ \citenamefont {Flambaum}}]{DzuDerFla12a}%
  \BibitemOpen
  \bibfield  {author} {\bibinfo {author} {\bibfnamefont {V.~A.}\ \bibnamefont
  {Dzuba}}, \bibinfo {author} {\bibfnamefont {A.}~\bibnamefont {Derevianko}}, \
  and\ \bibinfo {author} {\bibfnamefont {V.~V.}\ \bibnamefont {Flambaum}},\
  }\href {\doibase 10.1103/PhysRevA.86.054501} {\bibfield  {journal} {\bibinfo
  {journal} {Phys. Rev. A}\ }\textbf {\bibinfo {volume} {86}},\ \bibinfo
  {pages} {054501} (\bibinfo {year} {2012}{\natexlab{a}})}\BibitemShut
  {NoStop}%
\bibitem [{\citenamefont {Dzuba}\ \emph
  {et~al.}(2012{\natexlab{b}})\citenamefont {Dzuba}, \citenamefont
  {Derevianko},\ and\ \citenamefont {Flambaum}}]{DzuDerFla12b}%
  \BibitemOpen
  \bibfield  {author} {\bibinfo {author} {\bibfnamefont {V.~A.}\ \bibnamefont
  {Dzuba}}, \bibinfo {author} {\bibfnamefont {A.}~\bibnamefont {Derevianko}}, \
  and\ \bibinfo {author} {\bibfnamefont {V.~V.}\ \bibnamefont {Flambaum}},\
  }\href {\doibase 10.1103/PhysRevA.86.054502} {\bibfield  {journal} {\bibinfo
  {journal} {Phys. Rev. A}\ }\textbf {\bibinfo {volume} {86}},\ \bibinfo
  {pages} {054502} (\bibinfo {year} {2012}{\natexlab{b}})}\BibitemShut
  {NoStop}%
\bibitem [{\citenamefont {Kozlov}\ \emph {et~al.}(2013)\citenamefont {Kozlov},
  \citenamefont {Dzuba},\ and\ \citenamefont {Flambaum}}]{KozDzuFla13}%
  \BibitemOpen
  \bibfield  {author} {\bibinfo {author} {\bibfnamefont {A.}~\bibnamefont
  {Kozlov}}, \bibinfo {author} {\bibfnamefont {V.~A.}\ \bibnamefont {Dzuba}}, \
  and\ \bibinfo {author} {\bibfnamefont {V.~V.}\ \bibnamefont {Flambaum}},\
  }\href {\doibase 10.1103/PhysRevA.88.062509} {\bibfield  {journal} {\bibinfo
  {journal} {Phys. Rev. A}\ }\textbf {\bibinfo {volume} {88}},\ \bibinfo
  {pages} {062509} (\bibinfo {year} {2013})}\BibitemShut {NoStop}%
\bibitem [{\citenamefont {Safronova}\ \emph
  {et~al.}(2014{\natexlab{a}})\citenamefont {Safronova}, \citenamefont {Dzuba},
  \citenamefont {Flambaum}, \citenamefont {Safronova}, \citenamefont {Porsev},\
  and\ \citenamefont {Kozlov}}]{SafDzuFla14a}%
  \BibitemOpen
  \bibfield  {author} {\bibinfo {author} {\bibfnamefont {M.~S.}\ \bibnamefont
  {Safronova}}, \bibinfo {author} {\bibfnamefont {V.~A.}\ \bibnamefont
  {Dzuba}}, \bibinfo {author} {\bibfnamefont {V.~V.}\ \bibnamefont {Flambaum}},
  \bibinfo {author} {\bibfnamefont {U.~I.}\ \bibnamefont {Safronova}}, \bibinfo
  {author} {\bibfnamefont {S.~G.}\ \bibnamefont {Porsev}}, \ and\ \bibinfo
  {author} {\bibfnamefont {M.~G.}\ \bibnamefont {Kozlov}},\ }\href {\doibase
  10.1103/PhysRevA.90.052509} {\bibfield  {journal} {\bibinfo  {journal} {Phys.
  Rev. A}\ }\textbf {\bibinfo {volume} {90}},\ \bibinfo {pages} {052509}
  (\bibinfo {year} {2014}{\natexlab{a}})}\BibitemShut {NoStop}%
\bibitem [{\citenamefont {Safronova}\ \emph
  {et~al.}(2014{\natexlab{b}})\citenamefont {Safronova}, \citenamefont {Dzuba},
  \citenamefont {Flambaum}, \citenamefont {Safronova}, \citenamefont {Porsev},\
  and\ \citenamefont {Kozlov}}]{SafDzuFla14b}%
  \BibitemOpen
  \bibfield  {author} {\bibinfo {author} {\bibfnamefont {M.~S.}\ \bibnamefont
  {Safronova}}, \bibinfo {author} {\bibfnamefont {V.~A.}\ \bibnamefont
  {Dzuba}}, \bibinfo {author} {\bibfnamefont {V.~V.}\ \bibnamefont {Flambaum}},
  \bibinfo {author} {\bibfnamefont {U.~I.}\ \bibnamefont {Safronova}}, \bibinfo
  {author} {\bibfnamefont {S.~G.}\ \bibnamefont {Porsev}}, \ and\ \bibinfo
  {author} {\bibfnamefont {M.~G.}\ \bibnamefont {Kozlov}},\ }\href {\doibase
  10.1103/PhysRevA.90.042513} {\bibfield  {journal} {\bibinfo  {journal} {Phys.
  Rev. A}\ }\textbf {\bibinfo {volume} {90}},\ \bibinfo {pages} {042513}
  (\bibinfo {year} {2014}{\natexlab{b}})}\BibitemShut {NoStop}%
\bibitem [{\citenamefont {Safronova}\ \emph
  {et~al.}(2014{\natexlab{c}})\citenamefont {Safronova}, \citenamefont {Dzuba},
  \citenamefont {Flambaum}, \citenamefont {Safronova}, \citenamefont {Porsev},\
  and\ \citenamefont {Kozlov}}]{SafDzuFla14c}%
  \BibitemOpen
  \bibfield  {author} {\bibinfo {author} {\bibfnamefont {M.~S.}\ \bibnamefont
  {Safronova}}, \bibinfo {author} {\bibfnamefont {V.~A.}\ \bibnamefont
  {Dzuba}}, \bibinfo {author} {\bibfnamefont {V.~V.}\ \bibnamefont {Flambaum}},
  \bibinfo {author} {\bibfnamefont {U.~I.}\ \bibnamefont {Safronova}}, \bibinfo
  {author} {\bibfnamefont {S.~G.}\ \bibnamefont {Porsev}}, \ and\ \bibinfo
  {author} {\bibfnamefont {M.~G.}\ \bibnamefont {Kozlov}},\ }\href {\doibase
  10.1103/PhysRevLett.113.030801} {\bibfield  {journal} {\bibinfo  {journal}
  {Phys. Rev. Lett.}\ }\textbf {\bibinfo {volume} {113}},\ \bibinfo {pages}
  {030801} (\bibinfo {year} {2014}{\natexlab{c}})}\BibitemShut {NoStop}%
\bibitem [{\citenamefont {Yudin}\ \emph {et~al.}(2014)\citenamefont {Yudin},
  \citenamefont {Taichenachev},\ and\ \citenamefont
  {Derevianko}}]{YudTaiDer14}%
  \BibitemOpen
  \bibfield  {author} {\bibinfo {author} {\bibfnamefont {V.~I.}\ \bibnamefont
  {Yudin}}, \bibinfo {author} {\bibfnamefont {A.~V.}\ \bibnamefont
  {Taichenachev}}, \ and\ \bibinfo {author} {\bibfnamefont {A.}~\bibnamefont
  {Derevianko}},\ }\href {\doibase 10.1103/PhysRevLett.113.233003} {\bibfield
  {journal} {\bibinfo  {journal} {Phys. Rev. Lett.}\ }\textbf {\bibinfo
  {volume} {113}},\ \bibinfo {pages} {233003} (\bibinfo {year}
  {2014})}\BibitemShut {NoStop}%
\bibitem [{\citenamefont {Dzuba}\ \emph
  {et~al.}(2015{\natexlab{a}})\citenamefont {Dzuba}, \citenamefont {Flambaum},\
  and\ \citenamefont {Katori}}]{DzuFlaKat15}%
  \BibitemOpen
  \bibfield  {author} {\bibinfo {author} {\bibfnamefont {V.~A.}\ \bibnamefont
  {Dzuba}}, \bibinfo {author} {\bibfnamefont {V.~V.}\ \bibnamefont {Flambaum}},
  \ and\ \bibinfo {author} {\bibfnamefont {H.}~\bibnamefont {Katori}},\ }\href
  {\doibase 10.1103/PhysRevA.91.022119} {\bibfield  {journal} {\bibinfo
  {journal} {Phys. Rev. A}\ }\textbf {\bibinfo {volume} {91}},\ \bibinfo
  {pages} {022119} (\bibinfo {year} {2015}{\natexlab{a}})}\BibitemShut
  {NoStop}%
\bibitem [{\citenamefont {Dzuba}\ \emph
  {et~al.}(2015{\natexlab{b}})\citenamefont {Dzuba}, \citenamefont {Safronova},
  \citenamefont {Safronova},\ and\ \citenamefont {Flambaum}}]{DzuSafSaf15}%
  \BibitemOpen
  \bibfield  {author} {\bibinfo {author} {\bibfnamefont {V.~A.}\ \bibnamefont
  {Dzuba}}, \bibinfo {author} {\bibfnamefont {M.~S.}\ \bibnamefont
  {Safronova}}, \bibinfo {author} {\bibfnamefont {U.~I.}\ \bibnamefont
  {Safronova}}, \ and\ \bibinfo {author} {\bibfnamefont {V.~V.}\ \bibnamefont
  {Flambaum}},\ }\href {\doibase 10.1103/PhysRevA.92.060502} {\bibfield
  {journal} {\bibinfo  {journal} {Phys. Rev. A}\ }\textbf {\bibinfo {volume}
  {92}},\ \bibinfo {pages} {060502} (\bibinfo {year}
  {2015}{\natexlab{b}})}\BibitemShut {NoStop}%
\bibitem [{\citenamefont {Dzuba}\ and\ \citenamefont
  {Flambaum}(2015)}]{DzuFla15}%
  \BibitemOpen
  \bibfield  {author} {\bibinfo {author} {\bibfnamefont {V.~A.}\ \bibnamefont
  {Dzuba}}\ and\ \bibinfo {author} {\bibfnamefont {V.~V.}\ \bibnamefont
  {Flambaum}},\ }\href {\doibase 10.1007/s10751-015-1166-4} {\bibfield
  {journal} {\bibinfo  {journal} {Hyperfine Interact.}\ }\textbf {\bibinfo
  {volume} {236}},\ \bibinfo {pages} {79} (\bibinfo {year} {2015})}\BibitemShut
  {NoStop}%
\bibitem [{\citenamefont {Kozlov}\ \emph {et~al.}(2018)\citenamefont {Kozlov},
  \citenamefont {Safronova}, \citenamefont {Crespo L\'opez-Urrutia},\ and\
  \citenamefont {Schmidt}}]{KozSafCre18}%
  \BibitemOpen
  \bibfield  {author} {\bibinfo {author} {\bibfnamefont {M.~G.}\ \bibnamefont
  {Kozlov}}, \bibinfo {author} {\bibfnamefont {M.~S.}\ \bibnamefont
  {Safronova}}, \bibinfo {author} {\bibfnamefont {J.~R.}\ \bibnamefont {Crespo
  L\'opez-Urrutia}}, \ and\ \bibinfo {author} {\bibfnamefont {P.~O.}\
  \bibnamefont {Schmidt}},\ }\href {\doibase 10.1103/RevModPhys.90.045005}
  {\bibfield  {journal} {\bibinfo  {journal} {Rev. Mod. Phys.}\ }\textbf
  {\bibinfo {volume} {90}},\ \bibinfo {pages} {045005} (\bibinfo {year}
  {2018})}\BibitemShut {NoStop}%
\bibitem [{\citenamefont {Yu}\ and\ \citenamefont {Sahoo}(2019)}]{YuSah19}%
  \BibitemOpen
  \bibfield  {author} {\bibinfo {author} {\bibfnamefont {Y.-m.}\ \bibnamefont
  {Yu}}\ and\ \bibinfo {author} {\bibfnamefont {B.~K.}\ \bibnamefont {Sahoo}},\
  }\href {\doibase 10.1103/PhysRevA.99.022513} {\bibfield  {journal} {\bibinfo
  {journal} {Phys. Rev. A}\ }\textbf {\bibinfo {volume} {99}},\ \bibinfo
  {pages} {022513} (\bibinfo {year} {2019})}\BibitemShut {NoStop}%
\bibitem [{\citenamefont {Brewer}\ \emph {et~al.}(2013)\citenamefont {Brewer},
  \citenamefont {Guise},\ and\ \citenamefont {Tan}}]{BreGuiTan13}%
  \BibitemOpen
  \bibfield  {author} {\bibinfo {author} {\bibfnamefont {S.~M.}\ \bibnamefont
  {Brewer}}, \bibinfo {author} {\bibfnamefont {N.~D.}\ \bibnamefont {Guise}}, \
  and\ \bibinfo {author} {\bibfnamefont {J.~N.}\ \bibnamefont {Tan}},\ }\href
  {\doibase 10.1103/PhysRevA.88.063403} {\bibfield  {journal} {\bibinfo
  {journal} {Phys. Rev. A}\ }\textbf {\bibinfo {volume} {88}},\ \bibinfo
  {pages} {063403} (\bibinfo {year} {2013})}\BibitemShut {NoStop}%
\bibitem [{\citenamefont {Guise}\ \emph {et~al.}(2014)\citenamefont {Guise},
  \citenamefont {Tan}, \citenamefont {Brewer}, \citenamefont {Fischer},\ and\
  \citenamefont {J\"onsson}}]{GuiTanBre14}%
  \BibitemOpen
  \bibfield  {author} {\bibinfo {author} {\bibfnamefont {N.~D.}\ \bibnamefont
  {Guise}}, \bibinfo {author} {\bibfnamefont {J.~N.}\ \bibnamefont {Tan}},
  \bibinfo {author} {\bibfnamefont {S.~M.}\ \bibnamefont {Brewer}}, \bibinfo
  {author} {\bibfnamefont {C.~F.}\ \bibnamefont {Fischer}}, \ and\ \bibinfo
  {author} {\bibfnamefont {P.}~\bibnamefont {J\"onsson}},\ }\href {\doibase
  10.1103/PhysRevA.89.040502} {\bibfield  {journal} {\bibinfo  {journal} {Phys.
  Rev. A}\ }\textbf {\bibinfo {volume} {89}},\ \bibinfo {pages} {040502}
  (\bibinfo {year} {2014})}\BibitemShut {NoStop}%
\bibitem [{\citenamefont {Windberger}\ \emph {et~al.}(2015)\citenamefont
  {Windberger}, \citenamefont {Crespo L\'opez-Urrutia}, \citenamefont {Bekker},
  \citenamefont {Oreshkina}, \citenamefont {Berengut}, \citenamefont {Bock},
  \citenamefont {Borschevsky}, \citenamefont {Dzuba}, \citenamefont {Eliav},
  \citenamefont {Harman}, \citenamefont {Kaldor}, \citenamefont {Kaul},
  \citenamefont {Safronova}, \citenamefont {Flambaum}, \citenamefont {Keitel},
  \citenamefont {Schmidt}, \citenamefont {Ullrich},\ and\ \citenamefont
  {Versolato}}]{WinCreBek15}%
  \BibitemOpen
  \bibfield  {author} {\bibinfo {author} {\bibfnamefont {A.}~\bibnamefont
  {Windberger}}, \bibinfo {author} {\bibfnamefont {J.~R.}\ \bibnamefont {Crespo
  L\'opez-Urrutia}}, \bibinfo {author} {\bibfnamefont {H.}~\bibnamefont
  {Bekker}}, \bibinfo {author} {\bibfnamefont {N.~S.}\ \bibnamefont
  {Oreshkina}}, \bibinfo {author} {\bibfnamefont {J.~C.}\ \bibnamefont
  {Berengut}}, \bibinfo {author} {\bibfnamefont {V.}~\bibnamefont {Bock}},
  \bibinfo {author} {\bibfnamefont {A.}~\bibnamefont {Borschevsky}}, \bibinfo
  {author} {\bibfnamefont {V.~A.}\ \bibnamefont {Dzuba}}, \bibinfo {author}
  {\bibfnamefont {E.}~\bibnamefont {Eliav}}, \bibinfo {author} {\bibfnamefont
  {Z.}~\bibnamefont {Harman}}, \bibinfo {author} {\bibfnamefont
  {U.}~\bibnamefont {Kaldor}}, \bibinfo {author} {\bibfnamefont
  {S.}~\bibnamefont {Kaul}}, \bibinfo {author} {\bibfnamefont {U.~I.}\
  \bibnamefont {Safronova}}, \bibinfo {author} {\bibfnamefont {V.~V.}\
  \bibnamefont {Flambaum}}, \bibinfo {author} {\bibfnamefont {C.~H.}\
  \bibnamefont {Keitel}}, \bibinfo {author} {\bibfnamefont {P.~O.}\
  \bibnamefont {Schmidt}}, \bibinfo {author} {\bibfnamefont {J.}~\bibnamefont
  {Ullrich}}, \ and\ \bibinfo {author} {\bibfnamefont {O.~O.}\ \bibnamefont
  {Versolato}},\ }\href {\doibase 10.1103/PhysRevLett.114.150801} {\bibfield
  {journal} {\bibinfo  {journal} {Phys. Rev. Lett.}\ }\textbf {\bibinfo
  {volume} {114}},\ \bibinfo {pages} {150801} (\bibinfo {year}
  {2015})}\BibitemShut {NoStop}%
\bibitem [{\citenamefont {Schm{\"o}ger}\ \emph {et~al.}(2015)\citenamefont
  {Schm{\"o}ger}, \citenamefont {Versolato}, \citenamefont {Schwarz},
  \citenamefont {Kohnen}, \citenamefont {Windberger}, \citenamefont {Piest},
  \citenamefont {Feuchtenbeiner}, \citenamefont {Pedregosa-Gutierrez},
  \citenamefont {Leopold}, \citenamefont {Micke}, \citenamefont {Hansen},
  \citenamefont {Baumann}, \citenamefont {Drewsen}, \citenamefont {Ullrich},
  \citenamefont {Schmidt},\ and\ \citenamefont
  {L{\'o}pez-Urrutia}}]{SchVerSch15}%
  \BibitemOpen
  \bibfield  {author} {\bibinfo {author} {\bibfnamefont {L.}~\bibnamefont
  {Schm{\"o}ger}}, \bibinfo {author} {\bibfnamefont {O.~O.}\ \bibnamefont
  {Versolato}}, \bibinfo {author} {\bibfnamefont {M.}~\bibnamefont {Schwarz}},
  \bibinfo {author} {\bibfnamefont {M.}~\bibnamefont {Kohnen}}, \bibinfo
  {author} {\bibfnamefont {A.}~\bibnamefont {Windberger}}, \bibinfo {author}
  {\bibfnamefont {B.}~\bibnamefont {Piest}}, \bibinfo {author} {\bibfnamefont
  {S.}~\bibnamefont {Feuchtenbeiner}}, \bibinfo {author} {\bibfnamefont
  {J.}~\bibnamefont {Pedregosa-Gutierrez}}, \bibinfo {author} {\bibfnamefont
  {T.}~\bibnamefont {Leopold}}, \bibinfo {author} {\bibfnamefont
  {P.}~\bibnamefont {Micke}}, \bibinfo {author} {\bibfnamefont {A.~K.}\
  \bibnamefont {Hansen}}, \bibinfo {author} {\bibfnamefont {T.~M.}\
  \bibnamefont {Baumann}}, \bibinfo {author} {\bibfnamefont {M.}~\bibnamefont
  {Drewsen}}, \bibinfo {author} {\bibfnamefont {J.}~\bibnamefont {Ullrich}},
  \bibinfo {author} {\bibfnamefont {P.~O.}\ \bibnamefont {Schmidt}}, \ and\
  \bibinfo {author} {\bibfnamefont {J.~R.~C.}\ \bibnamefont
  {L{\'o}pez-Urrutia}},\ }\href {\doibase 10.1126/science.aaa2960} {\bibfield
  {journal} {\bibinfo  {journal} {Science}\ }\textbf {\bibinfo {volume}
  {347}},\ \bibinfo {pages} {1233} (\bibinfo {year} {2015})}\BibitemShut
  {NoStop}%
\bibitem [{\citenamefont {Leopold}\ \emph {et~al.}(2019)\citenamefont
  {Leopold}, \citenamefont {King}, \citenamefont {Micke}, \citenamefont
  {Bautista-Salvador}, \citenamefont {Heip}, \citenamefont {Ospelkaus},
  \citenamefont {López-Urrutia},\ and\ \citenamefont {Schmidt}}]{LeoKinMic19}%
  \BibitemOpen
  \bibfield  {author} {\bibinfo {author} {\bibfnamefont {T.}~\bibnamefont
  {Leopold}}, \bibinfo {author} {\bibfnamefont {S.~A.}\ \bibnamefont {King}},
  \bibinfo {author} {\bibfnamefont {P.}~\bibnamefont {Micke}}, \bibinfo
  {author} {\bibfnamefont {A.}~\bibnamefont {Bautista-Salvador}}, \bibinfo
  {author} {\bibfnamefont {J.~C.}\ \bibnamefont {Heip}}, \bibinfo {author}
  {\bibfnamefont {C.}~\bibnamefont {Ospelkaus}}, \bibinfo {author}
  {\bibfnamefont {J.~R.~C.}\ \bibnamefont {López-Urrutia}}, \ and\ \bibinfo
  {author} {\bibfnamefont {P.~O.}\ \bibnamefont {Schmidt}},\ }\href {\doibase
  10.1063/1.5100594} {\bibfield  {journal} {\bibinfo  {journal} {Rev. Sci.
  Instrum.}\ }\textbf {\bibinfo {volume} {90}},\ \bibinfo {pages} {073201}
  (\bibinfo {year} {2019})}\BibitemShut {NoStop}%
\bibitem [{\citenamefont {Bekker}\ \emph {et~al.}(2019)\citenamefont {Bekker},
  \citenamefont {Borschevsky}, \citenamefont {Harman}, \citenamefont {Keitel},
  \citenamefont {Pfeifer}, \citenamefont {Schmidt}, \citenamefont
  {López-Urrutia},\ and\ \citenamefont {Berengut}}]{BekBorHar19}%
  \BibitemOpen
  \bibfield  {author} {\bibinfo {author} {\bibfnamefont {H.}~\bibnamefont
  {Bekker}}, \bibinfo {author} {\bibfnamefont {A.}~\bibnamefont {Borschevsky}},
  \bibinfo {author} {\bibfnamefont {Z.}~\bibnamefont {Harman}}, \bibinfo
  {author} {\bibfnamefont {C.~H.}\ \bibnamefont {Keitel}}, \bibinfo {author}
  {\bibfnamefont {T.}~\bibnamefont {Pfeifer}}, \bibinfo {author} {\bibfnamefont
  {P.~O.}\ \bibnamefont {Schmidt}}, \bibinfo {author} {\bibfnamefont
  {J.~R.~C.}\ \bibnamefont {López-Urrutia}}, \ and\ \bibinfo {author}
  {\bibfnamefont {J.~C.}\ \bibnamefont {Berengut}},\ }\href {\doibase
  10.1038/s41467-019-13406-9} {\bibfield  {journal} {\bibinfo  {journal} {Nat.
  Commun.}\ }\textbf {\bibinfo {volume} {10}},\ \bibinfo {pages} {5651}
  (\bibinfo {year} {2019})}\BibitemShut {NoStop}%
\bibitem [{\citenamefont {Micke}\ \emph {et~al.}(2020)\citenamefont {Micke},
  \citenamefont {Leopold}, \citenamefont {King}, \citenamefont {Benkler},
  \citenamefont {Spiess}, \citenamefont {Schm\"{o}ger}, \citenamefont
  {Schwarz}, \citenamefont {L\'{o}pez-Urrutia},\ and\ \citenamefont
  {Schmidt}}]{MicLeoKin20}%
  \BibitemOpen
  \bibfield  {author} {\bibinfo {author} {\bibfnamefont {P.}~\bibnamefont
  {Micke}}, \bibinfo {author} {\bibfnamefont {T.}~\bibnamefont {Leopold}},
  \bibinfo {author} {\bibfnamefont {S.~A.}\ \bibnamefont {King}}, \bibinfo
  {author} {\bibfnamefont {E.}~\bibnamefont {Benkler}}, \bibinfo {author}
  {\bibfnamefont {L.~J.}\ \bibnamefont {Spiess}}, \bibinfo {author}
  {\bibfnamefont {L.}~\bibnamefont {Schm\"{o}ger}}, \bibinfo {author}
  {\bibfnamefont {M.}~\bibnamefont {Schwarz}}, \bibinfo {author} {\bibfnamefont
  {J.~R.~C.}\ \bibnamefont {L\'{o}pez-Urrutia}}, \ and\ \bibinfo {author}
  {\bibfnamefont {P.~O.}\ \bibnamefont {Schmidt}},\ }\href {\doibase
  10.1038/s41586-020-1959-8} {\bibfield  {journal} {\bibinfo  {journal}
  {Nature}\ }\textbf {\bibinfo {volume} {578}},\ \bibinfo {pages} {60}
  (\bibinfo {year} {2020})}\BibitemShut {NoStop}%
\bibitem [{\citenamefont {Itano}\ \emph {et~al.}(1993)\citenamefont {Itano},
  \citenamefont {Bergquist}, \citenamefont {Bollinger}, \citenamefont
  {Gilligan}, \citenamefont {Heinzen}, \citenamefont {Moore}, \citenamefont
  {Raizen},\ and\ \citenamefont {Wineland}}]{ItaBerBol93}%
  \BibitemOpen
  \bibfield  {author} {\bibinfo {author} {\bibfnamefont {W.~M.}\ \bibnamefont
  {Itano}}, \bibinfo {author} {\bibfnamefont {J.~C.}\ \bibnamefont
  {Bergquist}}, \bibinfo {author} {\bibfnamefont {J.~J.}\ \bibnamefont
  {Bollinger}}, \bibinfo {author} {\bibfnamefont {J.~M.}\ \bibnamefont
  {Gilligan}}, \bibinfo {author} {\bibfnamefont {D.~J.}\ \bibnamefont
  {Heinzen}}, \bibinfo {author} {\bibfnamefont {F.~L.}\ \bibnamefont {Moore}},
  \bibinfo {author} {\bibfnamefont {M.~G.}\ \bibnamefont {Raizen}}, \ and\
  \bibinfo {author} {\bibfnamefont {D.~J.}\ \bibnamefont {Wineland}},\ }\href
  {\doibase 10.1103/PhysRevA.47.3554} {\bibfield  {journal} {\bibinfo
  {journal} {Phys. Rev. A}\ }\textbf {\bibinfo {volume} {47}},\ \bibinfo
  {pages} {3554} (\bibinfo {year} {1993})}\BibitemShut {NoStop}%
\bibitem [{\citenamefont {Peik}\ \emph {et~al.}(2005)\citenamefont {Peik},
  \citenamefont {Schneider},\ and\ \citenamefont {Tamm}}]{PeiSchTam05}%
  \BibitemOpen
  \bibfield  {author} {\bibinfo {author} {\bibfnamefont {E.}~\bibnamefont
  {Peik}}, \bibinfo {author} {\bibfnamefont {T.}~\bibnamefont {Schneider}}, \
  and\ \bibinfo {author} {\bibfnamefont {C.}~\bibnamefont {Tamm}},\ }\href
  {\doibase 10.1088/0953-4075/39/1/012} {\bibfield  {journal} {\bibinfo
  {journal} {J. Phys. B}\ }\textbf {\bibinfo {volume} {39}},\ \bibinfo {pages}
  {145} (\bibinfo {year} {2005})}\BibitemShut {NoStop}%
\bibitem [{\citenamefont {Schioppo}\ \emph {et~al.}(2017)\citenamefont
  {Schioppo}, \citenamefont {Brown}, \citenamefont {McGrew}, \citenamefont
  {Hinkley}, \citenamefont {Fasano}, \citenamefont {Beloy}, \citenamefont
  {Yoon}, \citenamefont {Milani}, \citenamefont {Nicolodi}, \citenamefont
  {Sherman}, \citenamefont {Phillips}, \citenamefont {Oates},\ and\
  \citenamefont {Ludlow}}]{SchBroMcG17}%
  \BibitemOpen
  \bibfield  {author} {\bibinfo {author} {\bibfnamefont {M.}~\bibnamefont
  {Schioppo}}, \bibinfo {author} {\bibfnamefont {R.~C.}\ \bibnamefont {Brown}},
  \bibinfo {author} {\bibfnamefont {W.~F.}\ \bibnamefont {McGrew}}, \bibinfo
  {author} {\bibfnamefont {N.}~\bibnamefont {Hinkley}}, \bibinfo {author}
  {\bibfnamefont {R.~J.}\ \bibnamefont {Fasano}}, \bibinfo {author}
  {\bibfnamefont {K.}~\bibnamefont {Beloy}}, \bibinfo {author} {\bibfnamefont
  {T.~H.}\ \bibnamefont {Yoon}}, \bibinfo {author} {\bibfnamefont
  {G.}~\bibnamefont {Milani}}, \bibinfo {author} {\bibfnamefont
  {D.}~\bibnamefont {Nicolodi}}, \bibinfo {author} {\bibfnamefont {J.~A.}\
  \bibnamefont {Sherman}}, \bibinfo {author} {\bibfnamefont {N.~B.}\
  \bibnamefont {Phillips}}, \bibinfo {author} {\bibfnamefont {C.~W.}\
  \bibnamefont {Oates}}, \ and\ \bibinfo {author} {\bibfnamefont {A.~D.}\
  \bibnamefont {Ludlow}},\ }\href {\doibase 10.1038/nphoton.2016.231}
  {\bibfield  {journal} {\bibinfo  {journal} {Nat. Photon.}\ }\textbf {\bibinfo
  {volume} {11}},\ \bibinfo {pages} {48} (\bibinfo {year} {2017})}\BibitemShut
  {NoStop}%
\bibitem [{\citenamefont {Oelker}\ \emph {et~al.}(2019)\citenamefont {Oelker},
  \citenamefont {Hutson}, \citenamefont {Kennedy}, \citenamefont {Sonderhouse},
  \citenamefont {Bothwell}, \citenamefont {Goban}, \citenamefont {Kedar},
  \citenamefont {Sanner}, \citenamefont {Robinson}, \citenamefont {Marti},
  \citenamefont {Matei}, \citenamefont {Legero}, \citenamefont {Giunta},
  \citenamefont {Holzwarth}, \citenamefont {Riehle}, \citenamefont {Sterr},\
  and\ \citenamefont {Ye}}]{OelHutKen19}%
  \BibitemOpen
  \bibfield  {author} {\bibinfo {author} {\bibfnamefont {E.}~\bibnamefont
  {Oelker}}, \bibinfo {author} {\bibfnamefont {R.~B.}\ \bibnamefont {Hutson}},
  \bibinfo {author} {\bibfnamefont {C.~J.}\ \bibnamefont {Kennedy}}, \bibinfo
  {author} {\bibfnamefont {L.}~\bibnamefont {Sonderhouse}}, \bibinfo {author}
  {\bibfnamefont {T.}~\bibnamefont {Bothwell}}, \bibinfo {author}
  {\bibfnamefont {A.}~\bibnamefont {Goban}}, \bibinfo {author} {\bibfnamefont
  {D.}~\bibnamefont {Kedar}}, \bibinfo {author} {\bibfnamefont
  {C.}~\bibnamefont {Sanner}}, \bibinfo {author} {\bibfnamefont {J.~M.}\
  \bibnamefont {Robinson}}, \bibinfo {author} {\bibfnamefont {G.~E.}\
  \bibnamefont {Marti}}, \bibinfo {author} {\bibfnamefont {D.~G.}\ \bibnamefont
  {Matei}}, \bibinfo {author} {\bibfnamefont {T.}~\bibnamefont {Legero}},
  \bibinfo {author} {\bibfnamefont {M.}~\bibnamefont {Giunta}}, \bibinfo
  {author} {\bibfnamefont {R.}~\bibnamefont {Holzwarth}}, \bibinfo {author}
  {\bibfnamefont {F.}~\bibnamefont {Riehle}}, \bibinfo {author} {\bibfnamefont
  {U.}~\bibnamefont {Sterr}}, \ and\ \bibinfo {author} {\bibfnamefont
  {J.}~\bibnamefont {Ye}},\ }\href {\doibase 10.1038/s41566-019-0493-4}
  {\bibfield  {journal} {\bibinfo  {journal} {Nat. Photon.}\ }\textbf {\bibinfo
  {volume} {13}},\ \bibinfo {pages} {714} (\bibinfo {year} {2019})}\BibitemShut
  {NoStop}%
\bibitem [{\citenamefont {Peik}\ and\ \citenamefont {Tamm}(2003)}]{PeiTam03}%
  \BibitemOpen
  \bibfield  {author} {\bibinfo {author} {\bibfnamefont {E.}~\bibnamefont
  {Peik}}\ and\ \bibinfo {author} {\bibfnamefont {C.}~\bibnamefont {Tamm}},\
  }\href {\doibase 10.1209/epl/i2003-00210-x} {\bibfield  {journal} {\bibinfo
  {journal} {Europhys. Lett.}\ }\textbf {\bibinfo {volume} {61}},\ \bibinfo
  {pages} {181} (\bibinfo {year} {2003})}\BibitemShut {NoStop}%
\bibitem [{\citenamefont {Seiferle}\ \emph {et~al.}(2019)\citenamefont
  {Seiferle}, \citenamefont {von~der Wense}, \citenamefont {Bilous},
  \citenamefont {Amersdorffer}, \citenamefont {Christoph~Lemell}, \citenamefont
  {Stellmer}, \citenamefont {Schumm}, \citenamefont {D\"{u}llmann},
  \citenamefont {P\'{a}lffy},\ and\ \citenamefont {Thirolf}}]{SeiWenBil19}%
  \BibitemOpen
  \bibfield  {author} {\bibinfo {author} {\bibfnamefont {B.}~\bibnamefont
  {Seiferle}}, \bibinfo {author} {\bibfnamefont {L.}~\bibnamefont {von~der
  Wense}}, \bibinfo {author} {\bibfnamefont {P.~V.}\ \bibnamefont {Bilous}},
  \bibinfo {author} {\bibfnamefont {I.}~\bibnamefont {Amersdorffer}}, \bibinfo
  {author} {\bibfnamefont {F.~L.}\ \bibnamefont {Christoph~Lemell}}, \bibinfo
  {author} {\bibfnamefont {S.}~\bibnamefont {Stellmer}}, \bibinfo {author}
  {\bibfnamefont {T.}~\bibnamefont {Schumm}}, \bibinfo {author} {\bibfnamefont
  {C.~E.}\ \bibnamefont {D\"{u}llmann}}, \bibinfo {author} {\bibfnamefont
  {A.}~\bibnamefont {P\'{a}lffy}}, \ and\ \bibinfo {author} {\bibfnamefont
  {P.~G.}\ \bibnamefont {Thirolf}},\ }\href {\doibase
  10.1038/s41586-019-1533-4} {\bibfield  {journal} {\bibinfo  {journal}
  {Nature}\ }\textbf {\bibinfo {volume} {573}},\ \bibinfo {pages} {243}
  (\bibinfo {year} {2019})}\BibitemShut {NoStop}%
\bibitem [{\citenamefont {Yudin}\ \emph {et~al.}(2010)\citenamefont {Yudin},
  \citenamefont {Taichenachev}, \citenamefont {Oates}, \citenamefont {Barber},
  \citenamefont {Lemke}, \citenamefont {Ludlow}, \citenamefont {Sterr},
  \citenamefont {Lisdat},\ and\ \citenamefont {Riehle}}]{YudTaiOat10}%
  \BibitemOpen
  \bibfield  {author} {\bibinfo {author} {\bibfnamefont {V.~I.}\ \bibnamefont
  {Yudin}}, \bibinfo {author} {\bibfnamefont {A.~V.}\ \bibnamefont
  {Taichenachev}}, \bibinfo {author} {\bibfnamefont {C.~W.}\ \bibnamefont
  {Oates}}, \bibinfo {author} {\bibfnamefont {Z.~W.}\ \bibnamefont {Barber}},
  \bibinfo {author} {\bibfnamefont {N.~D.}\ \bibnamefont {Lemke}}, \bibinfo
  {author} {\bibfnamefont {A.~D.}\ \bibnamefont {Ludlow}}, \bibinfo {author}
  {\bibfnamefont {U.}~\bibnamefont {Sterr}}, \bibinfo {author} {\bibfnamefont
  {C.}~\bibnamefont {Lisdat}}, \ and\ \bibinfo {author} {\bibfnamefont
  {F.}~\bibnamefont {Riehle}},\ }\href {\doibase 10.1103/PhysRevA.82.011804}
  {\bibfield  {journal} {\bibinfo  {journal} {Phys. Rev. A}\ }\textbf {\bibinfo
  {volume} {82}},\ \bibinfo {pages} {011804} (\bibinfo {year}
  {2010})}\BibitemShut {NoStop}%
\bibitem [{\citenamefont {Sanner}\ \emph {et~al.}(2018)\citenamefont {Sanner},
  \citenamefont {Huntemann}, \citenamefont {Lange}, \citenamefont {Tamm},\ and\
  \citenamefont {Peik}}]{SanHunLan18}%
  \BibitemOpen
  \bibfield  {author} {\bibinfo {author} {\bibfnamefont {C.}~\bibnamefont
  {Sanner}}, \bibinfo {author} {\bibfnamefont {N.}~\bibnamefont {Huntemann}},
  \bibinfo {author} {\bibfnamefont {R.}~\bibnamefont {Lange}}, \bibinfo
  {author} {\bibfnamefont {C.}~\bibnamefont {Tamm}}, \ and\ \bibinfo {author}
  {\bibfnamefont {E.}~\bibnamefont {Peik}},\ }\href {\doibase
  10.1103/PhysRevLett.120.053602} {\bibfield  {journal} {\bibinfo  {journal}
  {Phys. Rev. Lett.}\ }\textbf {\bibinfo {volume} {120}},\ \bibinfo {pages}
  {053602} (\bibinfo {year} {2018})}\BibitemShut {NoStop}%
\bibitem [{\citenamefont {Beloy}(2018)}]{Bel18}%
  \BibitemOpen
  \bibfield  {author} {\bibinfo {author} {\bibfnamefont {K.}~\bibnamefont
  {Beloy}},\ }\href {\doibase 10.1103/PhysRevA.97.031406} {\bibfield  {journal}
  {\bibinfo  {journal} {Phys. Rev. A}\ }\textbf {\bibinfo {volume} {97}},\
  \bibinfo {pages} {031406} (\bibinfo {year} {2018})}\BibitemShut {NoStop}%
\bibitem [{\citenamefont {Matei}\ \emph {et~al.}(2017)\citenamefont {Matei},
  \citenamefont {Legero}, \citenamefont {H\"afner}, \citenamefont {Grebing},
  \citenamefont {Weyrich}, \citenamefont {Zhang}, \citenamefont {Sonderhouse},
  \citenamefont {Robinson}, \citenamefont {Ye}, \citenamefont {Riehle},\ and\
  \citenamefont {Sterr}}]{MatLegHaf17}%
  \BibitemOpen
  \bibfield  {author} {\bibinfo {author} {\bibfnamefont {D.~G.}\ \bibnamefont
  {Matei}}, \bibinfo {author} {\bibfnamefont {T.}~\bibnamefont {Legero}},
  \bibinfo {author} {\bibfnamefont {S.}~\bibnamefont {H\"afner}}, \bibinfo
  {author} {\bibfnamefont {C.}~\bibnamefont {Grebing}}, \bibinfo {author}
  {\bibfnamefont {R.}~\bibnamefont {Weyrich}}, \bibinfo {author} {\bibfnamefont
  {W.}~\bibnamefont {Zhang}}, \bibinfo {author} {\bibfnamefont
  {L.}~\bibnamefont {Sonderhouse}}, \bibinfo {author} {\bibfnamefont {J.~M.}\
  \bibnamefont {Robinson}}, \bibinfo {author} {\bibfnamefont {J.}~\bibnamefont
  {Ye}}, \bibinfo {author} {\bibfnamefont {F.}~\bibnamefont {Riehle}}, \ and\
  \bibinfo {author} {\bibfnamefont {U.}~\bibnamefont {Sterr}},\ }\href
  {\doibase 10.1103/PhysRevLett.118.263202} {\bibfield  {journal} {\bibinfo
  {journal} {Phys. Rev. Lett.}\ }\textbf {\bibinfo {volume} {118}},\ \bibinfo
  {pages} {263202} (\bibinfo {year} {2017})}\BibitemShut {NoStop}%
\bibitem [{\citenamefont {Robinson}\ \emph {et~al.}(2019)\citenamefont
  {Robinson}, \citenamefont {Oelker}, \citenamefont {Milner}, \citenamefont
  {Zhang}, \citenamefont {Legero}, \citenamefont {Matei}, \citenamefont
  {Riehle}, \citenamefont {Sterr},\ and\ \citenamefont {Ye}}]{RobOelMil19}%
  \BibitemOpen
  \bibfield  {author} {\bibinfo {author} {\bibfnamefont {J.~M.}\ \bibnamefont
  {Robinson}}, \bibinfo {author} {\bibfnamefont {E.}~\bibnamefont {Oelker}},
  \bibinfo {author} {\bibfnamefont {W.~R.}\ \bibnamefont {Milner}}, \bibinfo
  {author} {\bibfnamefont {W.}~\bibnamefont {Zhang}}, \bibinfo {author}
  {\bibfnamefont {T.}~\bibnamefont {Legero}}, \bibinfo {author} {\bibfnamefont
  {D.~G.}\ \bibnamefont {Matei}}, \bibinfo {author} {\bibfnamefont
  {F.}~\bibnamefont {Riehle}}, \bibinfo {author} {\bibfnamefont
  {U.}~\bibnamefont {Sterr}}, \ and\ \bibinfo {author} {\bibfnamefont
  {J.}~\bibnamefont {Ye}},\ }\href {\doibase 10.1364/OPTICA.6.000240}
  {\bibfield  {journal} {\bibinfo  {journal} {Optica}\ }\textbf {\bibinfo
  {volume} {6}},\ \bibinfo {pages} {240} (\bibinfo {year} {2019})}\BibitemShut
  {NoStop}%
\bibitem [{\citenamefont {Reader}(1983)}]{Rea83}%
  \BibitemOpen
  \bibfield  {author} {\bibinfo {author} {\bibfnamefont {J.}~\bibnamefont
  {Reader}},\ }\href {\doibase 10.1364/JOSA.73.000349} {\bibfield  {journal}
  {\bibinfo  {journal} {J. Opt. Soc. Am.}\ }\textbf {\bibinfo {volume} {73}},\
  \bibinfo {pages} {349} (\bibinfo {year} {1983})}\BibitemShut {NoStop}%
\bibitem [{\citenamefont {Micke}\ \emph {et~al.}(2018)\citenamefont {Micke},
  \citenamefont {K\"uhn}, \citenamefont {Buchauer}, \citenamefont {Harries},
  \citenamefont {B\"ucking}, \citenamefont {Blaum}, \citenamefont {Cieluch},
  \citenamefont {Egl}, \citenamefont {Hollain}, \citenamefont {Kraemer},
  \citenamefont {Pfeifer}, \citenamefont {Schmidt}, \citenamefont
  {Sch\"ussler}, \citenamefont {Schweiger}, \citenamefont {St\"ohlker},
  \citenamefont {Sturm}, \citenamefont {Wolf}, \citenamefont {Bernitt},\ and\
  \citenamefont {Crespo L\'opez-Urrutia}}]{MickKuhBuc18}%
  \BibitemOpen
  \bibfield  {author} {\bibinfo {author} {\bibfnamefont {P.}~\bibnamefont
  {Micke}}, \bibinfo {author} {\bibfnamefont {S.}~\bibnamefont {K\"uhn}},
  \bibinfo {author} {\bibfnamefont {L.}~\bibnamefont {Buchauer}}, \bibinfo
  {author} {\bibfnamefont {J.~R.}\ \bibnamefont {Harries}}, \bibinfo {author}
  {\bibfnamefont {T.~M.}\ \bibnamefont {B\"ucking}}, \bibinfo {author}
  {\bibfnamefont {K.}~\bibnamefont {Blaum}}, \bibinfo {author} {\bibfnamefont
  {A.}~\bibnamefont {Cieluch}}, \bibinfo {author} {\bibfnamefont
  {A.}~\bibnamefont {Egl}}, \bibinfo {author} {\bibfnamefont {D.}~\bibnamefont
  {Hollain}}, \bibinfo {author} {\bibfnamefont {S.}~\bibnamefont {Kraemer}},
  \bibinfo {author} {\bibfnamefont {T.}~\bibnamefont {Pfeifer}}, \bibinfo
  {author} {\bibfnamefont {P.~O.}\ \bibnamefont {Schmidt}}, \bibinfo {author}
  {\bibfnamefont {R.~X.}\ \bibnamefont {Sch\"ussler}}, \bibinfo {author}
  {\bibfnamefont {C.}~\bibnamefont {Schweiger}}, \bibinfo {author}
  {\bibfnamefont {T.}~\bibnamefont {St\"ohlker}}, \bibinfo {author}
  {\bibfnamefont {S.}~\bibnamefont {Sturm}}, \bibinfo {author} {\bibfnamefont
  {R.~N.}\ \bibnamefont {Wolf}}, \bibinfo {author} {\bibfnamefont
  {S.}~\bibnamefont {Bernitt}}, \ and\ \bibinfo {author} {\bibfnamefont
  {J.~R.}\ \bibnamefont {Crespo L\'opez-Urrutia}},\ }\href {\doibase
  10.1063/1.5026961} {\bibfield  {journal} {\bibinfo  {journal} {Rev. Sci.
  Instrum.}\ }\textbf {\bibinfo {volume} {89}},\ \bibinfo {pages} {063109}
  (\bibinfo {year} {2018})}\BibitemShut {NoStop}%
\bibitem [{NIS()}]{NISTbasicASD}%
  \BibitemOpen
  \href {\doibase 10.18434/T4FW23} {}\bibinfo {note} {NIST Basic Atomic
  Spectroscopic Data,
  \url{https://physics.nist.gov/PhysRefData/Handbook/periodictable.htm}}\BibitemShut
  {NoStop}%
\bibitem [{\citenamefont {Beloy}\ \emph {et~al.}(2017)\citenamefont {Beloy},
  \citenamefont {Leibrandt},\ and\ \citenamefont {Itano}}]{BelLeiIta17}%
  \BibitemOpen
  \bibfield  {author} {\bibinfo {author} {\bibfnamefont {K.}~\bibnamefont
  {Beloy}}, \bibinfo {author} {\bibfnamefont {D.~R.}\ \bibnamefont
  {Leibrandt}}, \ and\ \bibinfo {author} {\bibfnamefont {W.~M.}\ \bibnamefont
  {Itano}},\ }\href {\doibase 10.1103/PhysRevA.95.043405} {\bibfield  {journal}
  {\bibinfo  {journal} {Phys. Rev. A}\ }\textbf {\bibinfo {volume} {95}},\
  \bibinfo {pages} {043405} (\bibinfo {year} {2017})}\BibitemShut {NoStop}%
\bibitem [{\citenamefont {Itano}(2000)}]{Ita00}%
  \BibitemOpen
  \bibfield  {author} {\bibinfo {author} {\bibfnamefont {W.~M.}\ \bibnamefont
  {Itano}},\ }\href {\doibase 10.6028/jres.105.065} {\bibfield  {journal}
  {\bibinfo  {journal} {J. Res. Natl. Inst. Stand. Technol.}\ }\textbf
  {\bibinfo {volume} {105}},\ \bibinfo {pages} {829} (\bibinfo {year}
  {2000})}\BibitemShut {NoStop}%
\bibitem [{\citenamefont {Dub\'e}\ \emph {et~al.}(2005)\citenamefont {Dub\'e},
  \citenamefont {Madej}, \citenamefont {Bernard}, \citenamefont {Marmet},
  \citenamefont {Boulanger},\ and\ \citenamefont {Cundy}}]{DubMadBer05}%
  \BibitemOpen
  \bibfield  {author} {\bibinfo {author} {\bibfnamefont {P.}~\bibnamefont
  {Dub\'e}}, \bibinfo {author} {\bibfnamefont {A.~A.}\ \bibnamefont {Madej}},
  \bibinfo {author} {\bibfnamefont {J.~E.}\ \bibnamefont {Bernard}}, \bibinfo
  {author} {\bibfnamefont {L.}~\bibnamefont {Marmet}}, \bibinfo {author}
  {\bibfnamefont {J.-S.}\ \bibnamefont {Boulanger}}, \ and\ \bibinfo {author}
  {\bibfnamefont {S.}~\bibnamefont {Cundy}},\ }\href {\doibase
  10.1103/PhysRevLett.95.033001} {\bibfield  {journal} {\bibinfo  {journal}
  {Phys. Rev. Lett.}\ }\textbf {\bibinfo {volume} {95}},\ \bibinfo {pages}
  {033001} (\bibinfo {year} {2005})}\BibitemShut {NoStop}%
\bibitem [{\citenamefont {Dzuba}\ \emph {et~al.}(1996)\citenamefont {Dzuba},
  \citenamefont {Flambaum},\ and\ \citenamefont {Kozlov}}]{DzuFlaKoz96}%
  \BibitemOpen
  \bibfield  {author} {\bibinfo {author} {\bibfnamefont {V.~A.}\ \bibnamefont
  {Dzuba}}, \bibinfo {author} {\bibfnamefont {V.~V.}\ \bibnamefont {Flambaum}},
  \ and\ \bibinfo {author} {\bibfnamefont {M.~G.}\ \bibnamefont {Kozlov}},\
  }\href {\doibase 10.1103/PhysRevA.54.3948} {\bibfield  {journal} {\bibinfo
  {journal} {Phys. Rev. A}\ }\textbf {\bibinfo {volume} {54}},\ \bibinfo
  {pages} {3948} (\bibinfo {year} {1996})}\BibitemShut {NoStop}%
\bibitem [{\citenamefont {Dzuba}(2005)}]{Dzu05}%
  \BibitemOpen
  \bibfield  {author} {\bibinfo {author} {\bibfnamefont {V.~A.}\ \bibnamefont
  {Dzuba}},\ }\href {\doibase 10.1103/PhysRevA.71.032512} {\bibfield  {journal}
  {\bibinfo  {journal} {Phys. Rev. A}\ }\textbf {\bibinfo {volume} {71}},\
  \bibinfo {pages} {032512} (\bibinfo {year} {2005})}\BibitemShut {NoStop}%
\bibitem [{\citenamefont {Dzuba}\ \emph {et~al.}(2017)\citenamefont {Dzuba},
  \citenamefont {Berengut}, \citenamefont {Harabati},\ and\ \citenamefont
  {Flambaum}}]{DzuBerHar17}%
  \BibitemOpen
  \bibfield  {author} {\bibinfo {author} {\bibfnamefont {V.~A.}\ \bibnamefont
  {Dzuba}}, \bibinfo {author} {\bibfnamefont {J.~C.}\ \bibnamefont {Berengut}},
  \bibinfo {author} {\bibfnamefont {C.}~\bibnamefont {Harabati}}, \ and\
  \bibinfo {author} {\bibfnamefont {V.~V.}\ \bibnamefont {Flambaum}},\ }\href
  {\doibase 10.1103/PhysRevA.95.012503} {\bibfield  {journal} {\bibinfo
  {journal} {Phys. Rev. A}\ }\textbf {\bibinfo {volume} {95}},\ \bibinfo
  {pages} {012503} (\bibinfo {year} {2017})}\BibitemShut {NoStop}%
\bibitem [{\citenamefont {Dzuba}\ \emph {et~al.}(1987)\citenamefont {Dzuba},
  \citenamefont {Flambaum}, \citenamefont {Silvestrov},\ and\ \citenamefont
  {Sushkov}}]{DzuFlaSil87}%
  \BibitemOpen
  \bibfield  {author} {\bibinfo {author} {\bibfnamefont {V.~A.}\ \bibnamefont
  {Dzuba}}, \bibinfo {author} {\bibfnamefont {V.~V.}\ \bibnamefont {Flambaum}},
  \bibinfo {author} {\bibfnamefont {P.~G.}\ \bibnamefont {Silvestrov}}, \ and\
  \bibinfo {author} {\bibfnamefont {O.~P.}\ \bibnamefont {Sushkov}},\ }\href
  {\doibase 10.1088/0022-3700/20/7/009} {\bibfield  {journal} {\bibinfo
  {journal} {J. Phys. B}\ }\textbf {\bibinfo {volume} {20}},\ \bibinfo {pages}
  {1399} (\bibinfo {year} {1987})}\BibitemShut {NoStop}%
\bibitem [{\citenamefont {Dub\'e}\ \emph {et~al.}(2014)\citenamefont {Dub\'e},
  \citenamefont {Madej}, \citenamefont {Tibbo},\ and\ \citenamefont
  {Bernard}}]{DubMadTib14}%
  \BibitemOpen
  \bibfield  {author} {\bibinfo {author} {\bibfnamefont {P.}~\bibnamefont
  {Dub\'e}}, \bibinfo {author} {\bibfnamefont {A.~A.}\ \bibnamefont {Madej}},
  \bibinfo {author} {\bibfnamefont {M.}~\bibnamefont {Tibbo}}, \ and\ \bibinfo
  {author} {\bibfnamefont {J.~E.}\ \bibnamefont {Bernard}},\ }\href {\doibase
  10.1103/PhysRevLett.112.173002} {\bibfield  {journal} {\bibinfo  {journal}
  {Phys. Rev. Lett.}\ }\textbf {\bibinfo {volume} {112}},\ \bibinfo {pages}
  {173002} (\bibinfo {year} {2014})}\BibitemShut {NoStop}%
\bibitem [{\citenamefont {Barrett}\ \emph {et~al.}(2019)\citenamefont
  {Barrett}, \citenamefont {Arnold},\ and\ \citenamefont
  {Safronova}}]{BarArnSaf19}%
  \BibitemOpen
  \bibfield  {author} {\bibinfo {author} {\bibfnamefont {M.~D.}\ \bibnamefont
  {Barrett}}, \bibinfo {author} {\bibfnamefont {K.~J.}\ \bibnamefont {Arnold}},
  \ and\ \bibinfo {author} {\bibfnamefont {M.~S.}\ \bibnamefont {Safronova}},\
  }\href {\doibase 10.1103/PhysRevA.100.043418} {\bibfield  {journal} {\bibinfo
   {journal} {Phys. Rev. A}\ }\textbf {\bibinfo {volume} {100}},\ \bibinfo
  {pages} {043418} (\bibinfo {year} {2019})}\BibitemShut {NoStop}%
\bibitem [{\citenamefont {Rosenband}\ \emph {et~al.}(2008)\citenamefont
  {Rosenband}, \citenamefont {Hume}, \citenamefont {Schmidt}, \citenamefont
  {Chou}, \citenamefont {Brusch}, \citenamefont {Lorini}, \citenamefont
  {Oskay}, \citenamefont {Drullinger}, \citenamefont {Fortier}, \citenamefont
  {Stalnaker}, \citenamefont {Diddams}, \citenamefont {Swann}, \citenamefont
  {Newbury}, \citenamefont {Itano}, \citenamefont {Wineland},\ and\
  \citenamefont {Bergquist}}]{RosHumSch08}%
  \BibitemOpen
  \bibfield  {author} {\bibinfo {author} {\bibfnamefont {T.}~\bibnamefont
  {Rosenband}}, \bibinfo {author} {\bibfnamefont {D.~B.}\ \bibnamefont {Hume}},
  \bibinfo {author} {\bibfnamefont {P.~O.}\ \bibnamefont {Schmidt}}, \bibinfo
  {author} {\bibfnamefont {C.~W.}\ \bibnamefont {Chou}}, \bibinfo {author}
  {\bibfnamefont {A.}~\bibnamefont {Brusch}}, \bibinfo {author} {\bibfnamefont
  {L.}~\bibnamefont {Lorini}}, \bibinfo {author} {\bibfnamefont {W.~H.}\
  \bibnamefont {Oskay}}, \bibinfo {author} {\bibfnamefont {R.~E.}\ \bibnamefont
  {Drullinger}}, \bibinfo {author} {\bibfnamefont {T.~M.}\ \bibnamefont
  {Fortier}}, \bibinfo {author} {\bibfnamefont {J.~E.}\ \bibnamefont
  {Stalnaker}}, \bibinfo {author} {\bibfnamefont {S.~A.}\ \bibnamefont
  {Diddams}}, \bibinfo {author} {\bibfnamefont {W.~C.}\ \bibnamefont {Swann}},
  \bibinfo {author} {\bibfnamefont {N.~R.}\ \bibnamefont {Newbury}}, \bibinfo
  {author} {\bibfnamefont {W.~M.}\ \bibnamefont {Itano}}, \bibinfo {author}
  {\bibfnamefont {D.~J.}\ \bibnamefont {Wineland}}, \ and\ \bibinfo {author}
  {\bibfnamefont {J.~C.}\ \bibnamefont {Bergquist}},\ }\href {\doibase
  10.1126/science.1154622} {\bibfield  {journal} {\bibinfo  {journal}
  {Science}\ }\textbf {\bibinfo {volume} {319}},\ \bibinfo {pages} {1808}
  (\bibinfo {year} {2008})}\BibitemShut {NoStop}%
\bibitem [{\citenamefont {Porsev}\ and\ \citenamefont
  {Derevianko}(2006)}]{PorDer06}%
  \BibitemOpen
  \bibfield  {author} {\bibinfo {author} {\bibfnamefont {S.~G.}\ \bibnamefont
  {Porsev}}\ and\ \bibinfo {author} {\bibfnamefont {A.}~\bibnamefont
  {Derevianko}},\ }\href {\doibase 10.1103/PhysRevA.74.020502} {\bibfield
  {journal} {\bibinfo  {journal} {Phys. Rev. A}\ }\textbf {\bibinfo {volume}
  {74}},\ \bibinfo {pages} {020502} (\bibinfo {year} {2006})}\BibitemShut
  {NoStop}%
\bibitem [{\citenamefont {Berkeland}\ \emph {et~al.}(1998)\citenamefont
  {Berkeland}, \citenamefont {Miller}, \citenamefont {Bergquist}, \citenamefont
  {Itano},\ and\ \citenamefont {Wineland}}]{BerMilBer98}%
  \BibitemOpen
  \bibfield  {author} {\bibinfo {author} {\bibfnamefont {D.~J.}\ \bibnamefont
  {Berkeland}}, \bibinfo {author} {\bibfnamefont {J.~D.}\ \bibnamefont
  {Miller}}, \bibinfo {author} {\bibfnamefont {J.~C.}\ \bibnamefont
  {Bergquist}}, \bibinfo {author} {\bibfnamefont {W.~M.}\ \bibnamefont
  {Itano}}, \ and\ \bibinfo {author} {\bibfnamefont {D.~J.}\ \bibnamefont
  {Wineland}},\ }\href {\doibase 10.1063/1.367318} {\bibfield  {journal}
  {\bibinfo  {journal} {J. Appl. Phys.}\ }\textbf {\bibinfo {volume} {83}},\
  \bibinfo {pages} {5025} (\bibinfo {year} {1998})}\BibitemShut {NoStop}%
\bibitem [{\citenamefont {Dub\'e}\ \emph {et~al.}(2013)\citenamefont {Dub\'e},
  \citenamefont {Madej}, \citenamefont {Zhou},\ and\ \citenamefont
  {Bernard}}]{DubMadZho13}%
  \BibitemOpen
  \bibfield  {author} {\bibinfo {author} {\bibfnamefont {P.}~\bibnamefont
  {Dub\'e}}, \bibinfo {author} {\bibfnamefont {A.~A.}\ \bibnamefont {Madej}},
  \bibinfo {author} {\bibfnamefont {Z.}~\bibnamefont {Zhou}}, \ and\ \bibinfo
  {author} {\bibfnamefont {J.~E.}\ \bibnamefont {Bernard}},\ }\href {\doibase
  10.1103/PhysRevA.87.023806} {\bibfield  {journal} {\bibinfo  {journal} {Phys.
  Rev. A}\ }\textbf {\bibinfo {volume} {87}},\ \bibinfo {pages} {023806}
  (\bibinfo {year} {2013})}\BibitemShut {NoStop}%
\bibitem [{\citenamefont {Huang}\ \emph {et~al.}(2019)\citenamefont {Huang},
  \citenamefont {Guan}, \citenamefont {Zeng}, \citenamefont {Tang},\ and\
  \citenamefont {Gao}}]{HuaGuaZen19}%
  \BibitemOpen
  \bibfield  {author} {\bibinfo {author} {\bibfnamefont {Y.}~\bibnamefont
  {Huang}}, \bibinfo {author} {\bibfnamefont {H.}~\bibnamefont {Guan}},
  \bibinfo {author} {\bibfnamefont {M.}~\bibnamefont {Zeng}}, \bibinfo {author}
  {\bibfnamefont {L.}~\bibnamefont {Tang}}, \ and\ \bibinfo {author}
  {\bibfnamefont {K.}~\bibnamefont {Gao}},\ }\href {\doibase
  10.1103/PhysRevA.99.011401} {\bibfield  {journal} {\bibinfo  {journal} {Phys.
  Rev. A}\ }\textbf {\bibinfo {volume} {99}},\ \bibinfo {pages} {011401}
  (\bibinfo {year} {2019})}\BibitemShut {NoStop}%
\bibitem [{\citenamefont {Arnold}\ \emph {et~al.}(2018)\citenamefont {Arnold},
  \citenamefont {Kaewuam}, \citenamefont {Roy}, \citenamefont {Tan},\ and\
  \citenamefont {Barrett}}]{ArnKaeRoy18}%
  \BibitemOpen
  \bibfield  {author} {\bibinfo {author} {\bibfnamefont {K.~J.}\ \bibnamefont
  {Arnold}}, \bibinfo {author} {\bibfnamefont {R.}~\bibnamefont {Kaewuam}},
  \bibinfo {author} {\bibfnamefont {A.}~\bibnamefont {Roy}}, \bibinfo {author}
  {\bibfnamefont {T.~R.}\ \bibnamefont {Tan}}, \ and\ \bibinfo {author}
  {\bibfnamefont {M.~D.}\ \bibnamefont {Barrett}},\ }\href {\doibase
  10.1038/s41467-018-04079-x} {\bibfield  {journal} {\bibinfo  {journal} {Nat.
  Commun.}\ }\textbf {\bibinfo {volume} {9}},\ \bibinfo {pages} {1650}
  (\bibinfo {year} {2018})}\BibitemShut {NoStop}%
\bibitem [{\citenamefont {Hannig}\ \emph {et~al.}(2019)\citenamefont {Hannig},
  \citenamefont {Pelzer}, \citenamefont {Scharnhorst}, \citenamefont {Kramer},
  \citenamefont {Stepanova}, \citenamefont {Xu}, \citenamefont {Spethmann},
  \citenamefont {Leroux}, \citenamefont {Mehlst\"aubler},\ and\ \citenamefont
  {Schmidt}}]{HanPelSch19}%
  \BibitemOpen
  \bibfield  {author} {\bibinfo {author} {\bibfnamefont {S.}~\bibnamefont
  {Hannig}}, \bibinfo {author} {\bibfnamefont {L.}~\bibnamefont {Pelzer}},
  \bibinfo {author} {\bibfnamefont {N.}~\bibnamefont {Scharnhorst}}, \bibinfo
  {author} {\bibfnamefont {J.}~\bibnamefont {Kramer}}, \bibinfo {author}
  {\bibfnamefont {M.}~\bibnamefont {Stepanova}}, \bibinfo {author}
  {\bibfnamefont {Z.~T.}\ \bibnamefont {Xu}}, \bibinfo {author} {\bibfnamefont
  {N.}~\bibnamefont {Spethmann}}, \bibinfo {author} {\bibfnamefont {I.~D.}\
  \bibnamefont {Leroux}}, \bibinfo {author} {\bibfnamefont {T.~E.}\
  \bibnamefont {Mehlst\"aubler}}, \ and\ \bibinfo {author} {\bibfnamefont
  {P.~O.}\ \bibnamefont {Schmidt}},\ }\href {\doibase 10.1063/1.5090583}
  {\bibfield  {journal} {\bibinfo  {journal} {Rev. Sci. Instrum.}\ }\textbf
  {\bibinfo {volume} {90}},\ \bibinfo {pages} {053204} (\bibinfo {year}
  {2019})}\BibitemShut {NoStop}%
\bibitem [{\citenamefont {Wan}\ \emph {et~al.}(2015)\citenamefont {Wan},
  \citenamefont {Gebert}, \citenamefont {Wolf},\ and\ \citenamefont
  {Schmidt}}]{WanGebWol15}%
  \BibitemOpen
  \bibfield  {author} {\bibinfo {author} {\bibfnamefont {Y.}~\bibnamefont
  {Wan}}, \bibinfo {author} {\bibfnamefont {F.}~\bibnamefont {Gebert}},
  \bibinfo {author} {\bibfnamefont {F.}~\bibnamefont {Wolf}}, \ and\ \bibinfo
  {author} {\bibfnamefont {P.~O.}\ \bibnamefont {Schmidt}},\ }\href {\doibase
  10.1103/PhysRevA.91.043425} {\bibfield  {journal} {\bibinfo  {journal} {Phys.
  Rev. A}\ }\textbf {\bibinfo {volume} {91}},\ \bibinfo {pages} {043425}
  (\bibinfo {year} {2015})}\BibitemShut {NoStop}%
\bibitem [{\citenamefont {Lechner}\ \emph {et~al.}(2016)\citenamefont
  {Lechner}, \citenamefont {Maier}, \citenamefont {Hempel}, \citenamefont
  {Jurcevic}, \citenamefont {Lanyon}, \citenamefont {Monz}, \citenamefont
  {Brownnutt}, \citenamefont {Blatt},\ and\ \citenamefont
  {Roos}}]{LecMaiHem16}%
  \BibitemOpen
  \bibfield  {author} {\bibinfo {author} {\bibfnamefont {R.}~\bibnamefont
  {Lechner}}, \bibinfo {author} {\bibfnamefont {C.}~\bibnamefont {Maier}},
  \bibinfo {author} {\bibfnamefont {C.}~\bibnamefont {Hempel}}, \bibinfo
  {author} {\bibfnamefont {P.}~\bibnamefont {Jurcevic}}, \bibinfo {author}
  {\bibfnamefont {B.~P.}\ \bibnamefont {Lanyon}}, \bibinfo {author}
  {\bibfnamefont {T.}~\bibnamefont {Monz}}, \bibinfo {author} {\bibfnamefont
  {M.}~\bibnamefont {Brownnutt}}, \bibinfo {author} {\bibfnamefont
  {R.}~\bibnamefont {Blatt}}, \ and\ \bibinfo {author} {\bibfnamefont {C.~F.}\
  \bibnamefont {Roos}},\ }\href {\doibase 10.1103/PhysRevA.93.053401}
  {\bibfield  {journal} {\bibinfo  {journal} {Phys. Rev. A}\ }\textbf {\bibinfo
  {volume} {93}},\ \bibinfo {pages} {053401} (\bibinfo {year}
  {2016})}\BibitemShut {NoStop}%
\bibitem [{\citenamefont {Chen}\ \emph {et~al.}(2017)\citenamefont {Chen},
  \citenamefont {Brewer}, \citenamefont {Chou}, \citenamefont {Wineland},
  \citenamefont {Leibrandt},\ and\ \citenamefont {Hume}}]{CheBreCho17}%
  \BibitemOpen
  \bibfield  {author} {\bibinfo {author} {\bibfnamefont {J.-S.}\ \bibnamefont
  {Chen}}, \bibinfo {author} {\bibfnamefont {S.~M.}\ \bibnamefont {Brewer}},
  \bibinfo {author} {\bibfnamefont {C.~W.}\ \bibnamefont {Chou}}, \bibinfo
  {author} {\bibfnamefont {D.~J.}\ \bibnamefont {Wineland}}, \bibinfo {author}
  {\bibfnamefont {D.~R.}\ \bibnamefont {Leibrandt}}, \ and\ \bibinfo {author}
  {\bibfnamefont {D.~B.}\ \bibnamefont {Hume}},\ }\href {\doibase
  10.1103/PhysRevLett.118.053002} {\bibfield  {journal} {\bibinfo  {journal}
  {Phys. Rev. Lett.}\ }\textbf {\bibinfo {volume} {118}},\ \bibinfo {pages}
  {053002} (\bibinfo {year} {2017})}\BibitemShut {NoStop}%
\bibitem [{\citenamefont {Dzuba}\ \emph {et~al.}(2018)\citenamefont {Dzuba},
  \citenamefont {Flambaum},\ and\ \citenamefont {Schiller}}]{DzuFlaSch18}%
  \BibitemOpen
  \bibfield  {author} {\bibinfo {author} {\bibfnamefont {V.~A.}\ \bibnamefont
  {Dzuba}}, \bibinfo {author} {\bibfnamefont {V.~V.}\ \bibnamefont {Flambaum}},
  \ and\ \bibinfo {author} {\bibfnamefont {S.}~\bibnamefont {Schiller}},\
  }\href {\doibase 10.1103/PhysRevA.98.022501} {\bibfield  {journal} {\bibinfo
  {journal} {Phys. Rev. A}\ }\textbf {\bibinfo {volume} {98}},\ \bibinfo
  {pages} {022501} (\bibinfo {year} {2018})}\BibitemShut {NoStop}%
\bibitem [{\citenamefont {Derevianko}\ and\ \citenamefont
  {Pospelov}(2014)}]{DerPos14}%
  \BibitemOpen
  \bibfield  {author} {\bibinfo {author} {\bibfnamefont {A.}~\bibnamefont
  {Derevianko}}\ and\ \bibinfo {author} {\bibfnamefont {M.}~\bibnamefont
  {Pospelov}},\ }\href {\doibase 10.1038/nphys3137} {\bibfield  {journal}
  {\bibinfo  {journal} {Nat. Phys.}\ }\textbf {\bibinfo {volume} {10}},\
  \bibinfo {pages} {933} (\bibinfo {year} {2014})}\BibitemShut {NoStop}%
\bibitem [{\citenamefont {Arvanitaki}\ \emph {et~al.}(2015)\citenamefont
  {Arvanitaki}, \citenamefont {Huang},\ and\ \citenamefont
  {Van~Tilburg}}]{ArvHuaVan15}%
  \BibitemOpen
  \bibfield  {author} {\bibinfo {author} {\bibfnamefont {A.}~\bibnamefont
  {Arvanitaki}}, \bibinfo {author} {\bibfnamefont {J.}~\bibnamefont {Huang}}, \
  and\ \bibinfo {author} {\bibfnamefont {K.}~\bibnamefont {Van~Tilburg}},\
  }\href {\doibase 10.1103/PhysRevD.91.015015} {\bibfield  {journal} {\bibinfo
  {journal} {Phys. Rev. D}\ }\textbf {\bibinfo {volume} {91}},\ \bibinfo
  {pages} {015015} (\bibinfo {year} {2015})}\BibitemShut {NoStop}%
\bibitem [{\citenamefont {Van~Tilburg}\ \emph {et~al.}(2015)\citenamefont
  {Van~Tilburg}, \citenamefont {Leefer}, \citenamefont {Bougas},\ and\
  \citenamefont {Budker}}]{VanLeeBou15}%
  \BibitemOpen
  \bibfield  {author} {\bibinfo {author} {\bibfnamefont {K.}~\bibnamefont
  {Van~Tilburg}}, \bibinfo {author} {\bibfnamefont {N.}~\bibnamefont {Leefer}},
  \bibinfo {author} {\bibfnamefont {L.}~\bibnamefont {Bougas}}, \ and\ \bibinfo
  {author} {\bibfnamefont {D.}~\bibnamefont {Budker}},\ }\href {\doibase
  10.1103/PhysRevLett.115.011802} {\bibfield  {journal} {\bibinfo  {journal}
  {Phys. Rev. Lett.}\ }\textbf {\bibinfo {volume} {115}},\ \bibinfo {pages}
  {011802} (\bibinfo {year} {2015})}\BibitemShut {NoStop}%
\bibitem [{\citenamefont {Stadnik}\ and\ \citenamefont
  {Flambaum}(2015)}]{StaFla15}%
  \BibitemOpen
  \bibfield  {author} {\bibinfo {author} {\bibfnamefont {Y.~V.}\ \bibnamefont
  {Stadnik}}\ and\ \bibinfo {author} {\bibfnamefont {V.~V.}\ \bibnamefont
  {Flambaum}},\ }\href {\doibase 10.1103/PhysRevLett.115.201301} {\bibfield
  {journal} {\bibinfo  {journal} {Phys. Rev. Lett.}\ }\textbf {\bibinfo
  {volume} {115}},\ \bibinfo {pages} {201301} (\bibinfo {year}
  {2015})}\BibitemShut {NoStop}%
\end{thebibliography}

%

\end{document}